\documentclass[a4paper,12pt,openany]{article}
\usepackage[top=20truemm,bottom=20truemm,left=22truemm,right=22truemm]{geometry}
\usepackage[]{hyperref}
\usepackage{amsmath}
\usepackage{ascmac}
\usepackage{amssymb}
\usepackage{ulem}
\usepackage[]{graphicx}

\graphicspath{{./figs/}}

\title{Complexity growth for topological black holes by holographic method}
\author{Koichi Nagasaki}
\date{\today}

\begin{document}
\vspace{1cm}

\begin{center}
	{\LARGE Complexity growth for topological black holes by holographic method}\\
\vspace{2cm}
	{\large Koichi Nagasaki}\footnote{koichi.nagasaki24@gmail.com}\\
\vspace{1cm}
	{\small Department of Physics, Toho University,\\
	Address: 2-2-1 Miyama, Funabashi, Chiba, 274-8510, Japan}
\end{center}
\vspace{1.5cm}

\abstract{
We consider the growth of the action for black hole spacetime with a fundamental string.
Our interest is to find the difference of the behavior between black holes with three different topologies in the scenario of complexity-action conjecture.
These black holes have positive, negative and zero curvatures.
We would like to calculate the action growth of these systems with a probe fundamental string according to the complexity-action conjecture.
We find that for the case where the black holes have the toroidal horizon structure this probe string behaves very differently from the other two cases.
}

\tableofcontents

\section{Introduction}\label{sec:intro}
The AdS/CFT correspondence is an important concept in recent theoretical physics \cite{Maldacena:1997re,Witten:1998qj,Awad:2000aj}.
By means of this correspondence, the strong coupling region of gauge theories can be studied by means of gravity method with small coupling method, and vice versa.
It will help both for gravitational and gauge theories.
Then it is interesting to find a new example of such a correspondence. 

One interesting method of this correspondence is the relation between a fundamental string motion in AdS spacetime and the drag force \cite{Gubser:2006bz, NataAtmaja:2010hd}.
In these works they added a probe string on the AdS spacetime.
An edge of the string is interpreted as a test particle on the dual gauge theory which lives on the infinite boundary of the AdS spacetime. 
This is related to the energy loss of the particle in the quark gluon plasma \cite{Herzog:2006gh, CasalderreySolana:2006rq,Liu:2006ug,Fadafan:2008bq,Fadafan:2012qu,Atashi:2016cso}.

A motivation to consider such probe strings is that nonlocal objects are useful to find a new property of the AdS/CFT correspondence, for example in our past work \cite{Nagasaki:2011ue}, where we confirmed an example of the correspondence between gravity and the gauge sides through a non-local object.
Examples of nonlocal objects are Wilson loops, 't Hooft operators, and so on.
Especially the string we consider in this paper corresponds to the Wilson loop.
By the equation of motion of the string, we will find the embedding of the fundamental string in black hole spacetime and calculate the action.
As explained later, there is a conjecture which relates this action to an important quantity to shed the light on the black hole physics.

I would like to explain one more motivation to study non-local operators in black hole spacetime.
In theoretical physics, black holes are expected as the fastest scrambler to perform a kind of calculation \cite{Sekino:2008he,Coleman:1991ku,Dvali:2016lnb,Swingle:2016var}.
To quantify such a freedom, recent physics supposes a new quantity.
It is ``complexity" \cite{0034-4885-75-2-022001,2008arXiv0804.3401W,Arora:2009:CCM:1540612, Moore:2011:NC:2086753,Susskind:2014moa, Susskind:2014rva, Caputa:2017yrh, Hashimoto:2018bmb, Susskind:2018fmx, Susskind:2018pmk, Bhattacharyya:2018wym}.
It has a similar property to entropy as an increasing function of time \cite{Brown:2017jil, Carmi:2017jqz, Fan:2019mbp}.
Roughly speaking, it counts the number of gates which is needed to cause the quantum development to a given state.
In quantum field theory, complexity is studied recently and this definition is revealed in many works \cite{Vanchurin:2016met, Chapman:2017rqy, Jefferson:2017sdb, Couch:2017yil, Caputa:2017yrh, Bhattacharyya:2018bbv, Ali:2018fcz, Jiang:2018gft, Guo:2018kzl, Yang:2018nda, Hackl:2018ptj, Khan:2018rzm, Yang:2018tpo, Bhattacharyya:2018wym, Chapman:2019clq, Yang:2019gce, Jafari:2019qns, Yang:2019udi, Sinamuli:2019utz, Guo:2019vni}.
Geometrical approaches are also studied \cite{2005quant.ph2070N, 2006Sci311.1133N, 2007quant.ph1004D, Brown:2016wib, Abt:2017pmf, Bao:2017qmt, Czech:2017ryf, Chapman:2018hou, Doroudiani:2019llj, Lin:2019kpf, Jiang:2018nzg, Brown:2019whu, Bernamonti:2019zyy}. 
In this sense, complexity is thought to be a geodesic on the circuit space where gates live.
Since holographic complexity is known to have nonlocal properties \cite{Moosa:2017yvt, Abad:2017cgl, Moosa:2017yvt, Fu:2018kcp, Numasawa:2018grg}, the nonlocal object will be essential to study the properties of complexity.
Complexity of systems with nonlocal operators such as ``Wilson loops" are studied also in our past works \cite{Ageev:2014nva, Nagasaki:2017kqe, Nagasaki:2018csh}.

Related to the AdS/CFT correspondence, some interesting conjectures for complexity were proposed recently.
A reliable candidate is complexity-action (CA) conjecture \cite{Brown:2015bva, Brown:2015lvg, Alishahiha:2015rta, Chapman:2016hwi, Tao:2017fsy, Jiang:2019fpz, Jiang:2019pgc}.
Its modified version is reported in \cite{Fan:2018wnv}.
This conjecture asserts the equivalence between complexity and the action calculated in a subspace of the bulk region called ``Wheeler DeWitt patch" (WDW).
This is the region bounded by null surfaces anchored at the given boundary time.
Complexity is expected to be a tool for solving many problems of black holes which includes firewalls, information paradox and so on \cite{Preskill:1992tc, Giddings:1993vj, Russo:2005aw, Hayden:2007cs, Terno:2009cc, Susskind:2012rm, Almheiri:2012rt, Harlow:2013tf, Maldacena:2013xja, Stanford:2014jda, Susskind:2014jwa, Stoltenberg:2014pua, Mann:2015luq, Barbon:2015ria, Susskind:2015toa, Momeni:2016ekm, Couch:2016exn, Roberts:2016hpo, Polchinski:2016hrw, Zhao:2017iul, Marolf:2017jkr, Gan:2017qkz, Ge:2017rak, HosseiniMansoori:2017tsm, Zangeneh:2017tub}. 
By this motivation, this conjecture is studied in the various spacetime geometries in many works 
\cite{2000Natur4061047L,Maldacena:2013xja,Stanford:2014jda,Barbon:2015soa, Pan:2016ecg,Huang:2016fks,Momeni:2016ira,Cai:2016xho,Hashimoto:2017fga,Alishahiha:2017hwg, Reynolds:2017lwq,Qaemmaqami:2017lzs,Yang:2017nfn,Karar:2017org,Miao:2017quj, Ghodrati:2017roz,Guo:2017rul,Sebastiani:2017rxr,Cai:2017sjv,Wang:2017uiw,Swingle:2017zcd, Cano:2018aqi,Brown:2018bms,Ghaffarnejad:2018bsd,Chakraborty:2018dvi,Chapman:2018dem, HosseiniMansoori:2018gdu,Mahapatra:2018gig,Ovgun:2018jbm,Chapman:2018lsv, Fareghbal:2018ngr,Bamba:2018ouz,Auzzi:2018pbc,Jiang:2018pfk,Ghaffarnejad:2018prc, Feng:2018sqm,Jiang:2018tlu,Meng:2018vtl,An:2018xhv,Ling:2018xpc,Fan:2018xwf,Abt:2018ywl, Huang:2019ajv,Fan:2019aoj,Caginalp:2019fyt,Ling:2019ien,An:2019opz,Jiang:2019qea, Goto:2018iay,Auzzi:2019mah,Liu:2019smx,Braccia:2019xxi,Jiang:2019yzs,ELMOUMNI2020114837}.

\begin{figure}
\begin{center}
	\includegraphics[width=11cm]{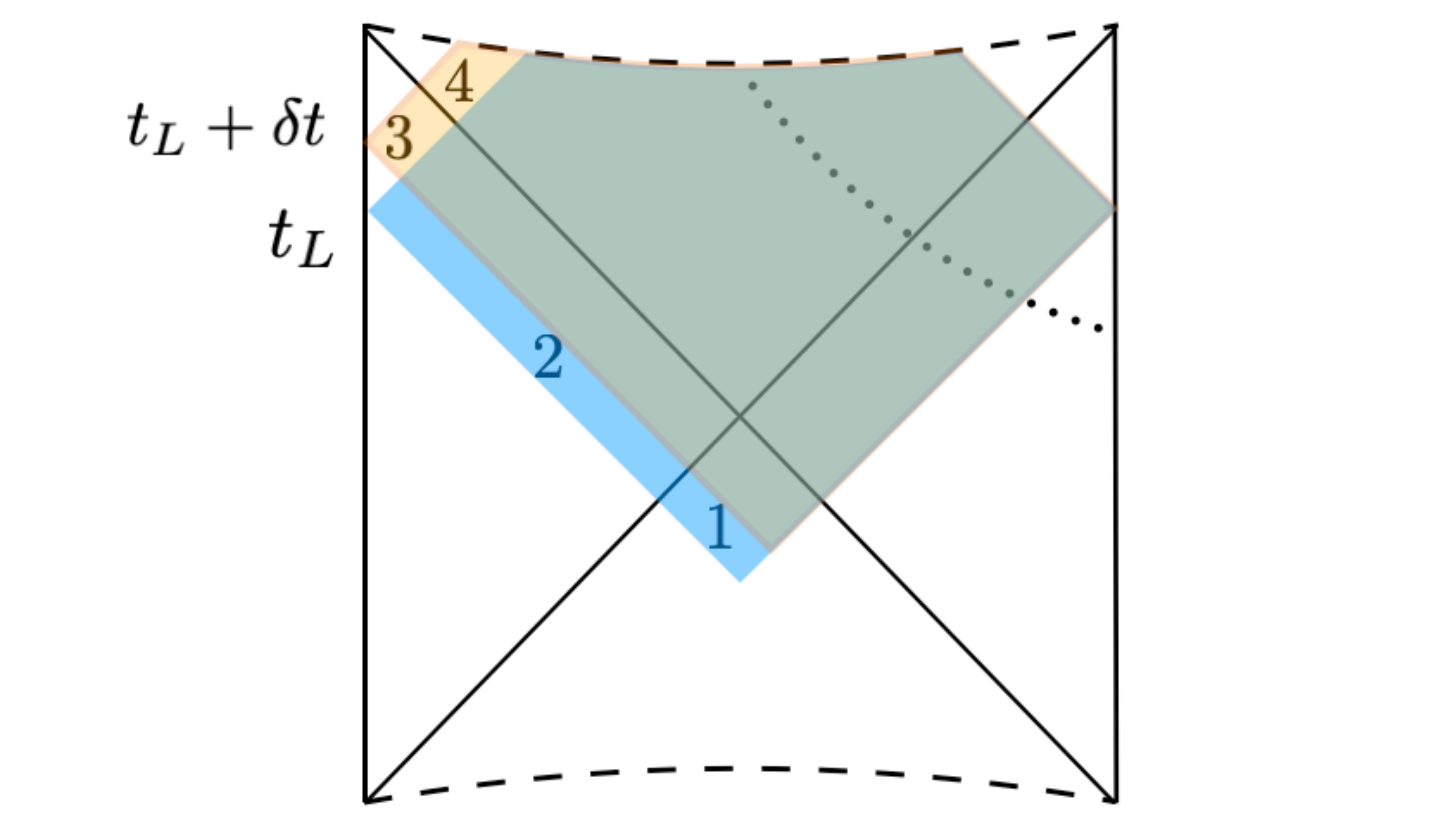}
	\caption{Penrose diagram of the AdS black holes:
	there are two CFTs on the left and the right sides.
	The WDW patch shaded in blue at time $t_L$ develops to another patch in time $\delta t$.
	The two diagonal lines represent the black hole horizon. 
	The dashed lines on the upper side and the lower side are singularity.
	The dotted line on the right side is a fundamental string.}
\label{fig:penrosediagram}
\end{center}
\end{figure}

We consider in this paper the two sided black holes which exist in AdS spacetime and this spacetime has two conformal field theories on the left and the right side of the AdS boundaries.
It is the most studied case for holographic method for black holes.
The Penrose diagram of these black holes is depicted in Figure \ref{fig:penrosediagram}.
In this figure the horizontal axis represents the radial direction and the vertical axis is the time direction.
In this figure the WDW patch anchored at left time $t_L$ (the blue region) develops to another patch by time $\delta t$.
The difference of these two region comes from the regions 1,2,3 and 4 in the figure.
As time go on, the region 1 and 2 disappear end the region 3 and 4 emerge.
As explained in Figure 2 of \cite{Brown:2015lvg}, the contribution from the region 1 is neglected at late times and the contributions from 2 and 3 cancel.
Therefore to find the growth of the action on the WDW we only have to integrate the Lagrangian in the region 4 which is bounded by the black hole horizon on the outside and the singularity.

For the transverse geometry of the Penrose diagram, we can take three different types. 
These are distinguished by the curvature and the metric is given in \eqref{eq:metric_3cases}.
We call these spacetime with different topologies ``topological black holes."
The topological black holes are objects which have interesting applications while hardly studied so far \cite{Vanzo:1997gw, Mann:1997iz, Birmingham:1998nr, Torii:2005nh, 10.1093/ptep/pty017, An:2018dbz}.
In this paper we would like to focus on these black holes and find the difference between these topologies.

This paper is organized as follows: 
In Section \ref{sec:statictopBH} we begin with static black hole spacetime with positive, negative and zero curvatures. 
We solve the equations of motion and find the action growth which is obtained by the integration over the WDW patch.
These do not have an interpretation in terms of a drag force because they do not evolve in time. 
However, we give this calculation for an example for introducing our method for simpler cases.
In Section \ref{sec:topKABH} we consider the rotating topological black holes.
In Section \ref{sec:Discussion} we conclude this paper by summarizing the results and some discussion.

\section{Static topological black holes}\label{sec:statictopBH}
In this section we consider a fundamental string on the static black hole spacetime.
Here the black holes have three different horizon structure whose curvatures are positive, negative or zero.
We consider the black hole space time which is asymptotically AdS$_{d+1}$.
The metric is described by \cite{Vanzo:1997gw, Mann:1997iz, Birmingham:1998nr, Chapman:2016hwi}
\begin{equation}\label{eq:static_adsmetric}
ds_{d+1}^2 = -f_k(r)dt^2 + \frac{dr^2}{f_k(r)} + r^2d\Sigma_{k,d-1}^2,
\end{equation}
where the metric function $f(r)$ is defined by
\begin{align}
f_k(r) = k - \frac{2m}{r^{d-2}} + \frac{r^2}{\ell^2}.
\end{align}
In the above parameter $k = \pm1,0$ distinguishes the topology of the horizon: these are the $(d-1)$-dimensional sphere, hyperboloid and torus.
The mass parameter and the AdS radius are denoted by $m$ and $\ell$, respectively.
In the following we appropriately rescale the radial coordinate $r$ and put $\ell = 1$. 
The angular part $d\Sigma_{k,d-1}$ is the metric on the ($d-1$)-dimensional Einstein space:
\begin{subequations}\label{eq:metric_3cases}
\begin{align}
&k=1&&
\text{sphere }(\mathbb S)&&:&
d\Sigma_{+1,d-1}^2 &= d\theta^2 + \sin^2\theta\; d\Omega_{d-2}^2,&
 \theta&\in[0,\pi],\\
&k=-1&&
\text{hyperbolic }(\mathbb H)&&:&
d\Sigma_{-1,d-1}^2 &= d\theta^2 + \sinh^2\theta\; d\Omega_{d-2}^2,&
 \theta&\in[0,\infty),\\
&k=0&&
\text{torus }(\mathbb T)&&:&
d\Sigma_{0,d-1}^2 &= d\theta^2 + \theta^2d\Omega_{d-2}^2,&
 \theta&\in[0,1].
\end{align}
\end{subequations}
where for the toroidal case coordinate $\theta$ is periodic.
\subsection{Action calculation}\label{sec:statictopBH_action} 
We consider a static string fixed at $\theta=\theta_0$ on the boundary.
By changing variables, $y = 1/r$, the metric is
\begin{align}
ds_{d+1}^2 
&= \frac1{y^2}\Big(-g_k(y)dt^2 + \frac{dy^2}{g_k(y)} + d\Sigma_{k,d-1}^2\Big),\\
g_k(y) &= y^2f_k(1/y) = 1 + ky^2 - 2my^d.
\end{align}
%
In the previous works \cite{Nagasaki:2017kqe,Nagasaki:2018csh}, we considered a string with constant angular velocity around the equator of the sphere. 
However on the hyperbolic space or torus geometry, there is no such a stationary motion.

Thus we consider here the case where the string is static.
This string is not moving then there do not exist an interpretation for the drag force. 
However in the next section we introduce the angular momentum of the black holes.
In this situation the string can occur the drag force.
In this section we then treat the static string and black holes as a simple exercise before dealing with the complicated calculation for rotating black holes.
The string worldsheet $(\tau,\sigma)$ is embedded as 
\begin{equation}\label{eq:static_embedding}
t = \tau,\quad
y = \sigma,\quad
\theta = \theta(\sigma);\quad
\sigma\in[0,\infty).
\end{equation}
The induced metric is
\begin{equation}
ds_\text{ind}^2
= \frac1{\sigma^2}\Big(-g_k(\sigma)d\tau^2 
  + \Big(\frac1{g_k(\sigma)} + \theta'(\sigma)^2\Big)d\sigma^2\Big).
\end{equation}
Note that $d\Omega_{d-2}$ part in \eqref{eq:metric_3cases} vanished since the string does not have the angular momentum and the difference between three topologies appears via only the function $g_k$.

\begin{figure}[t]
	\begin{minipage}[]{0.5\linewidth}
	\includegraphics[width=\linewidth]{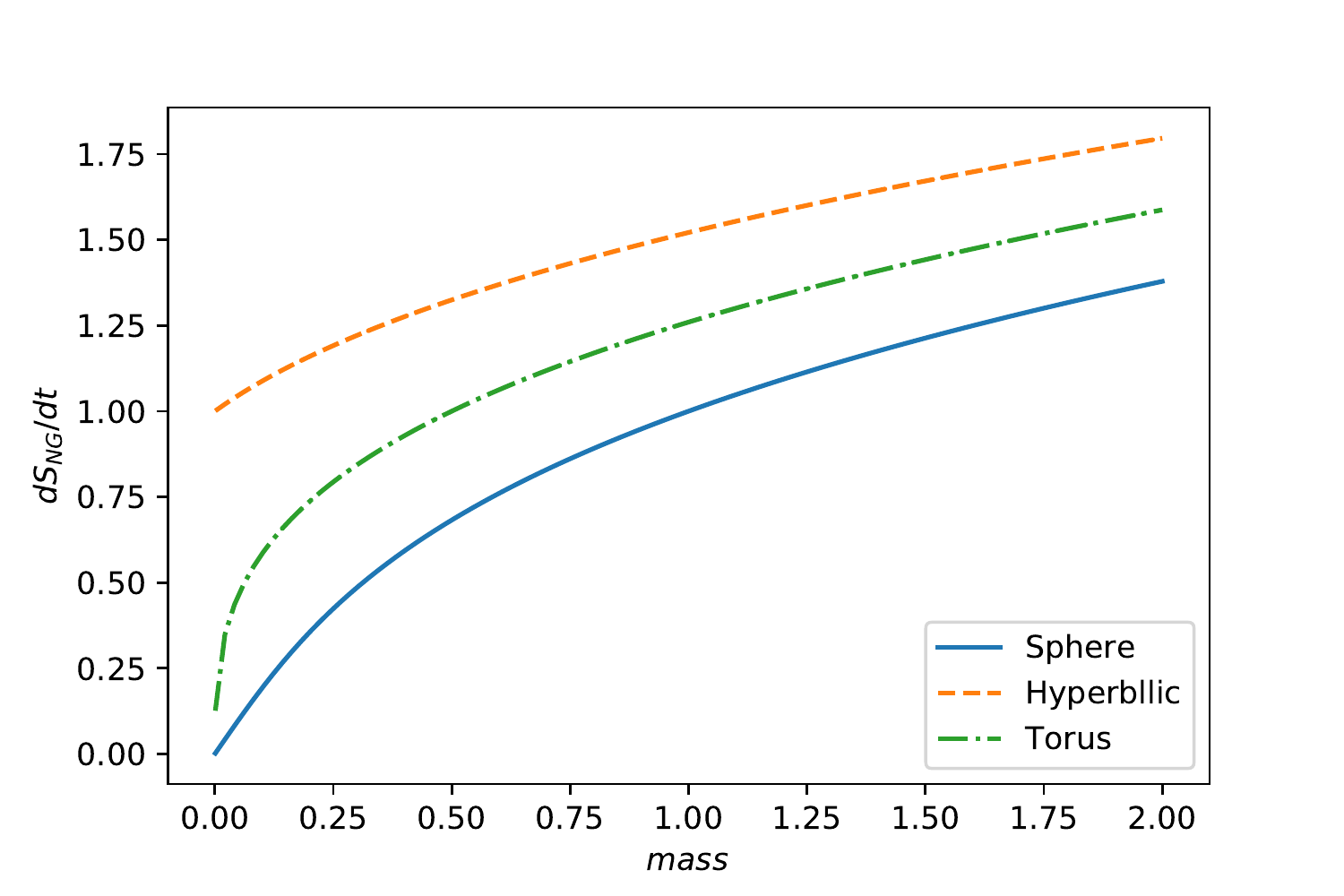}
	\caption{Mass and action growth: $d=3$}
	\label{fig:topBH_dim3_m}
	\end{minipage}
\hspace{0\linewidth}
	\begin{minipage}[]{0.5\linewidth}
	\includegraphics[width=\linewidth]{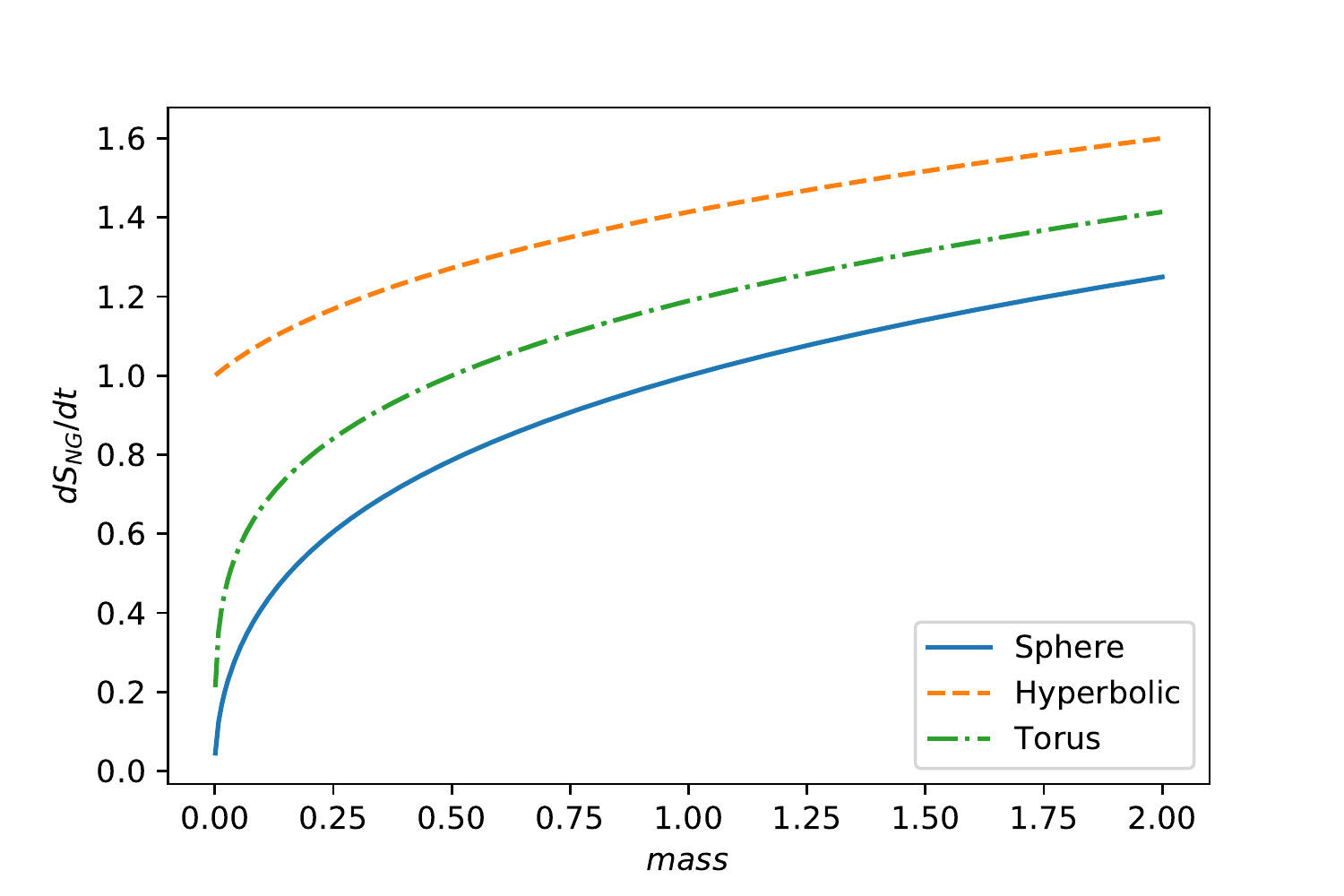}
	\caption{Mass and action growth: $d=4$}
	\label{fig:topBH_dim4_m}
	\end{minipage}
\end{figure}

The action consists of the gravitational terms for the black hole and the Nambu-Goto (NG) term for the probe string.
The effect of the string to the system is expressed by this NG action 
\begin{equation}
S_\text{NG} = T_\text{str}\int d\tau d\sigma\mathcal L(\sigma),
\end{equation}
where
\begin{equation}
\mathcal L(\sigma)
= \sqrt{-g_\text{ind}(\sigma)}
= \frac1{\sigma^2}\sqrt{1 + g_k(\sigma)\theta'(\sigma)^2}.
\end{equation}
The equation of motion for $\theta(\sigma)$ is 
\begin{equation}
\frac{d}{d\sigma}\bigg[\frac1{\sigma^2}\frac{g_k(\sigma)\theta'(\sigma)}{\sqrt{1+g_k(\sigma)\theta'(\sigma)^2}}\bigg] = 0.
\end{equation}
It can be integrated easily and gives 
\begin{equation}
\frac1{\sigma^2}\frac{g_k(\sigma)\theta'(\sigma)}{\sqrt{1+g_k(\sigma)\theta'(\sigma)^2}} = c,
\end{equation}
with integration constant $c$.
The above equation is solved for $\theta'(\sigma)$ as
\begin{equation}
\theta'(\sigma) 
 = \sqrt\frac{c^2\sigma^4}{g_k(\sigma)(g_k(\sigma) - c^2\sigma^4)}.
\end{equation}
This denominator of the inside of the square root becomes zero outside of the horizon where 
$g_k - c^2\sigma^4 = 0$ while the numerator is non-negative if $c>0$.
Then the only possibility that the string penetrates the horizon is $c=0$ and in this case $\theta = \text{const.}$
\paragraph{WDW action}
Let us calculate the NG action for Wheeler DeWitt patch in this case.
For $\theta'=0$ the Lagrangian is simplified as 
$\mathcal L = 1/\sigma^2$.
To find the growth rate of the action we integrate it inside of the horizon, namely, $y$ is integrated from $y_\text{h} := 1/r_\text{h}$ (horizon radius) to the infinity.
Then the NG action becomes
\begin{equation}
\frac1{T_\text{str}}\frac{dS_\text{NG}}{dt}
= \int_{y_\text{h}}^\infty\frac{d\sigma}{\sigma^2}
= \frac1{y_\text{h}} = r_\text{h}.
\end{equation}
It tells the action growth rate is determined by the location of the horizon $r_\text{h}$:
\begin{equation}\label{eq:statictopBH_horizon}
g_k(y_\text{h}) = 1 + ky_\text{h}^2 - 2my_\text{h}^{d} = 0.
\end{equation}
The mass dependence is plotted for 3 and 4 dimensions in Figure \ref{fig:topBH_dim3_m} and Figure \ref{fig:topBH_dim4_m}.
A remarkable point is that there is a non-trivial contribution even in massless case for hyperbolic black holes, as we can see from Equation \eqref{eq:statictopBH_horizon}: 
$0 = g_{-1}(y_\text{h})|_{m=0} = 1 - y_\text{h}^2$. 
There exists a non-trivial solution: $y_\text{h} = 1$.

\section{Rotating topological black holes}\label{sec:topKABH}
Rotating topological black holes are shown in \cite{Klemm:1997ea}.
The geometries of the ($d+1$)-dimensional rotating black hole with one rotational parameter $a$ are classified into three cases by the curvature $k$.
For $k=1$, the spacetime has the spherical horizon:
\begin{align}
ds_{\text{KA}\mathbb S}^2
&= -\frac{\Delta_r}{\rho^2}
  \Big(dt - \frac{a}{\Xi}\sin^2\theta d\phi\Big)^2
  + \frac{\rho^2}{\Delta_r}dr^2
  + \frac{\rho^2}{\Delta_\theta}d\theta^2\nonumber\\
&\qquad
  + \frac{\Delta_\theta\sin^2\theta}{\rho^2}
  \Big(adt - \frac{r^2+a^2}{\Xi}d\phi\Big)^2
  + r^2\cos^2\theta d\Omega_{1,d-2}^2,\\
\Delta_r &:= (r^2+a^2)(r^2+1) - \frac{2m}{r^{d-4}},\quad
\Delta_\theta := 1 - a^2\cos^2\theta,\nonumber\\
\rho^2 &:= r^2 + a^2\cos^2\theta,\quad
\Xi := 1 - a^2,\nonumber
\end{align}
where in subscript ``KA" denotes Kerr-AdS and ``$\mathbb {S}$" means sphere.
For the other two cases in the same way we denote them by 
$ds_{\text{KA}\mathbb H}^2$ and $ds_{\text{KA}\mathbb T}^2$.
For $k= -1$, the spacetime has the hyperbolic topology:
\begin{align}
ds_{\text{KA}\mathbb H}^2
&= -\frac{\Delta_r}{\rho^2}
  \Big(dt + \frac{a}{\Xi}\sinh^2\theta d\phi\Big)^2
  + \frac{\rho^2}{\Delta_r}dr^2
  + \frac{\rho^2}{\Delta_\theta}d\theta^2\nonumber\\
&\qquad
  + \frac{\Delta_\theta\sinh^2\theta}{\rho^2}
  \Big(adt - \frac{r^2+a^2}{\Xi}d\phi\Big)^2
  + r^2\cosh^2\theta d\Omega_{-1,d-2}^2,\\
\Delta_r &:= (r^2+a^2)(r^2-1) - \frac{2m}{r^{d-4}},\quad
\Delta_\theta := 1 + a^2\cosh^2\theta,\nonumber\\
\rho^2 &:= r^2 + a^2\cosh^2\theta,\quad
\Xi := 1 + a^2.\nonumber
\end{align}
For $k= 0$, the spacetime has the toroidal topology:
\begin{align}
ds_{\text{KA}\mathbb T}^2
&= - N^2dt^2 
  + \frac{\rho^2}{\Delta_r}dr^2
  + \frac{\rho^2}{\Delta_\theta}d\theta^2
  + \frac{\Sigma^2}{\rho^2}(d\phi - \omega dt)^2
  + r^2d\Omega_{0,d-2}^2,\\
\Delta_r &:= a^2 - \frac{2m}{r^{d-4}} + r^4,\quad
\Delta_\theta := 1 + a^2\theta^4,\nonumber\\
\rho^2 &:= r^2 + a^2\theta^2,\nonumber\\
\Sigma^2 &:= r^4\Delta_\theta - a^2\theta^4\Delta_r,\quad
\omega := a\frac{\theta^2\Delta_r + r^2\Delta_\theta}{\Sigma^2},\quad
N^2 := \frac{\rho^2\Delta_\theta\Delta_r}{\Sigma^2}.\nonumber
\end{align}
In each case, the parameter $a$ corresponds to the angular momentum per unit mass of the black hole as can been seen in $\omega$ in the torus case.
In $a\rightarrow0$ limit, these surely reproduce the static black holes as follows.
The factors of the metric become in this limit,
\begin{subequations}
\begin{align}
\text{Sphere}(\mathbb S)/\text{hyperbolic}(\mathbb H)&:
\Delta_r\rightarrow 
r^2(r^2\pm1) - \frac{2m}{r^{d-4}}
= r^2f(r),\quad
\Delta_\theta\rightarrow 1,\quad
\rho^2\rightarrow r^2,\nonumber\\
&\qquad
\Xi\rightarrow1,\\
\text{Torus}(\mathbb T)&:
\Delta_r\rightarrow
r^4 - \frac{2m}{r^{d-4}} = r^2f(r),\quad
\Delta_\theta\rightarrow1,\quad
\rho^2\rightarrow r^2,\label{eq:metric_func:SH}\nonumber\\
&\qquad
\Sigma^2\rightarrow r^4,\quad
\omega\rightarrow0,\quad
N^2\rightarrow\frac{r^4-2m/r^{d-4}}{r^2}
= f(r).
\end{align}
\end{subequations}
Then the time and radial parts become
$-f(r)dt^2 + dr^2/f(r)$ 
for all three cases and the angular part is 
\begin{subequations}\label{eq:metric_3cases_2}
\begin{align}
\mathbb S:&\quad
r^2d\theta^2 + r^2\sin^2\theta d\phi^2 
 + r^2\cos^2\theta d\Omega_{1,d-3}^2,\\
\mathbb H:&\quad
r^2d\theta^2 + r^2\sinh^2\theta d\phi^2
 + r^2\cosh^2\theta d\Omega_{-1,d-3}^2,\\
\mathbb T:&\quad
r^2d\theta^2 + r^2d\phi^2 + r^2d\Omega_{0,d-3}^2.
\end{align}
\end{subequations}
These are equal to the metric times radius $r$ shown in \eqref{eq:metric_3cases}, respectively.
For example, on three-sphere $\sum_{i=0}^{3}x_i^2 = 1$, if we choose spherical coordinates
\begin{equation}
x_0 = r\cos\psi,\;
x_1 = r\sin\psi\cos\theta,\;
x_2 = r\sin\psi\sin\theta\cos\varphi,\;
x_3 = r\sin\psi\sin\theta\sin\varphi,
\end{equation}
the metric is 
\begin{equation}
ds_{S^3}^2 = r^2(d\psi^2 + \sin^2\psi ds_{S^2}^2).
\end{equation}
On the other hand, if we choose the Hopf coordinates, 
\begin{equation}
x_0 = \cos\xi_1\sin\eta,\;
x_1 = \sin\xi_1\sin\eta,\;
x_2 = \cos\xi_2\cos\eta,\;
x_3 = \sin\xi_2\cos\eta,
\end{equation}
the same metric represented as 
\begin{equation}
ds_{S^3}^2 = d\eta^2 + \sin^2\eta\;d\xi_1^2 + \cos^2\eta\;d\xi_2^2.
\end{equation}
On the hyperbolic case, we can find the similar coordinates by changing $\sin\psi\rightarrow\sinh\psi$ and $\sinh\eta\rightarrow\sinh\eta$ from the sphere case. 
For the torus case, the metric \eqref{eq:metric_3cases} is the expression by the polar coordinate with radial direction $\theta$ and \eqref{eq:metric_3cases_2} is the expression by the ordinary flat metric. 
Then we confirm the equivalence between the previous metric \eqref{eq:metric_3cases} and the present case \eqref{eq:metric_3cases_2} and thus these Kerr-AdS metric recover the static cases \eqref{eq:static_adsmetric} in zero angular momentum limit.

The location of the horizon is depicted in the following figures (Figure \ref{fig:topBH_s_horizon}, Figure \ref{fig:topBH_h_horizon} and Figure \ref{fig:topBH_t_horizon}).
In all cases the horizon radii become small compared with the static black holes.

\begin{figure}[h]
	\begin{minipage}[t]{0.5\linewidth}
	\includegraphics[width=\linewidth]{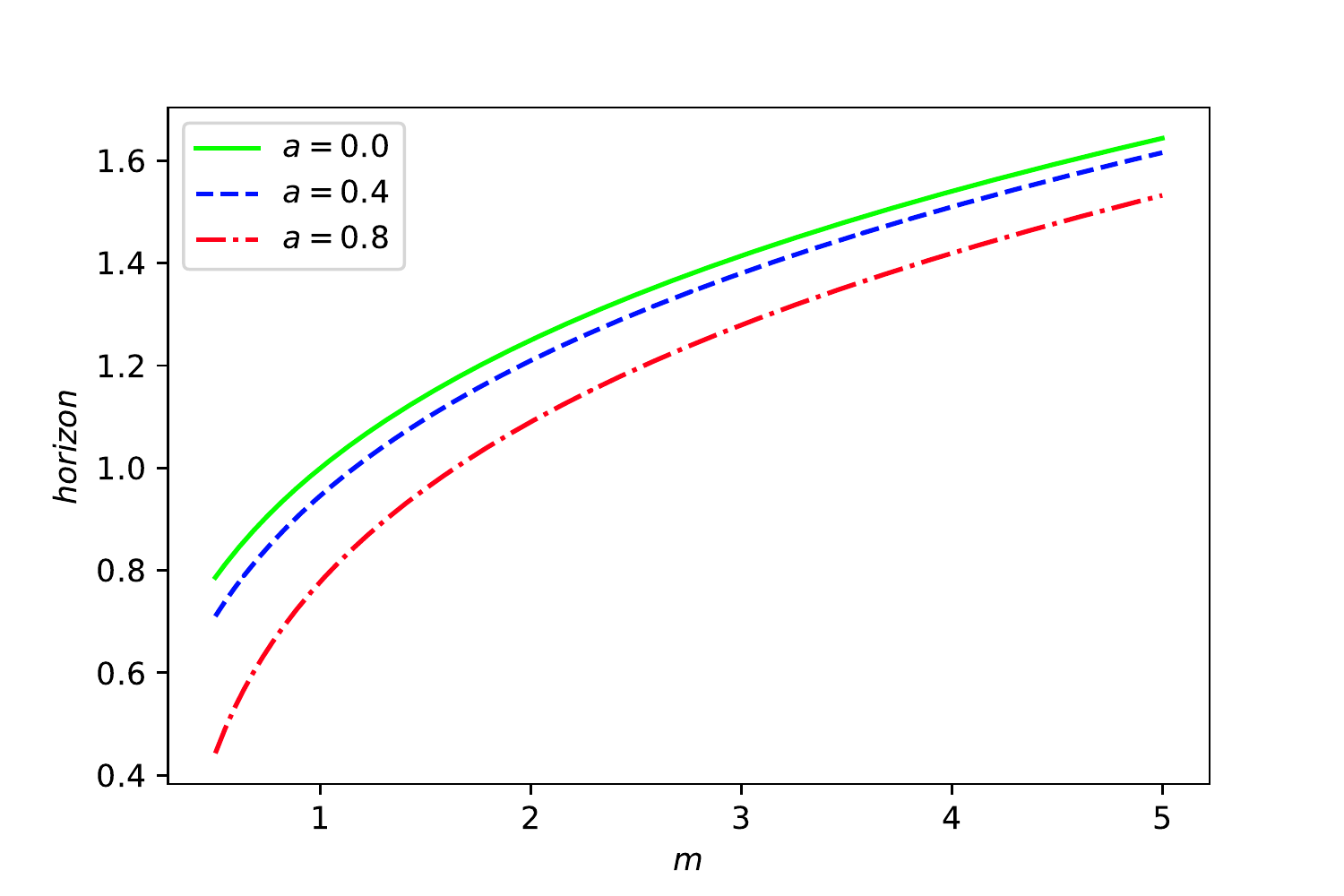}
	\caption{Mass and the horizon ($\mathbb S$): $d=4$}
	\label{fig:topBH_s_horizon}
	\end{minipage}
\hspace{0\linewidth}
	\begin{minipage}[t]{0.5\linewidth}
	\includegraphics[width=\linewidth]{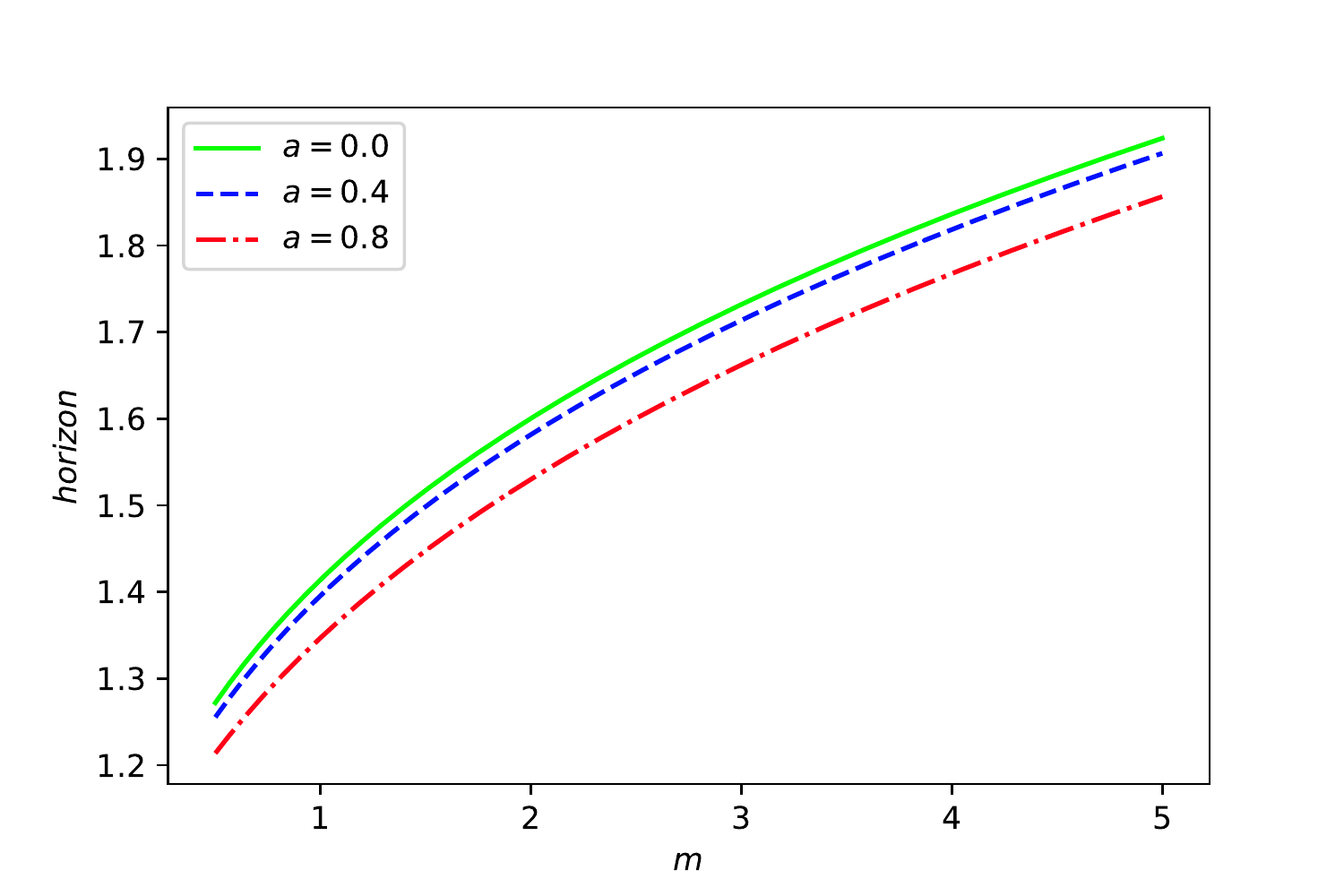}
	\caption{Mass and the horizon ($\mathbb H$): $d=4$}
	\label{fig:topBH_h_horizon}
	\end{minipage}
\hspace{0\linewidth}
	\begin{minipage}[t]{0.5\linewidth}
	\includegraphics[width=\linewidth]{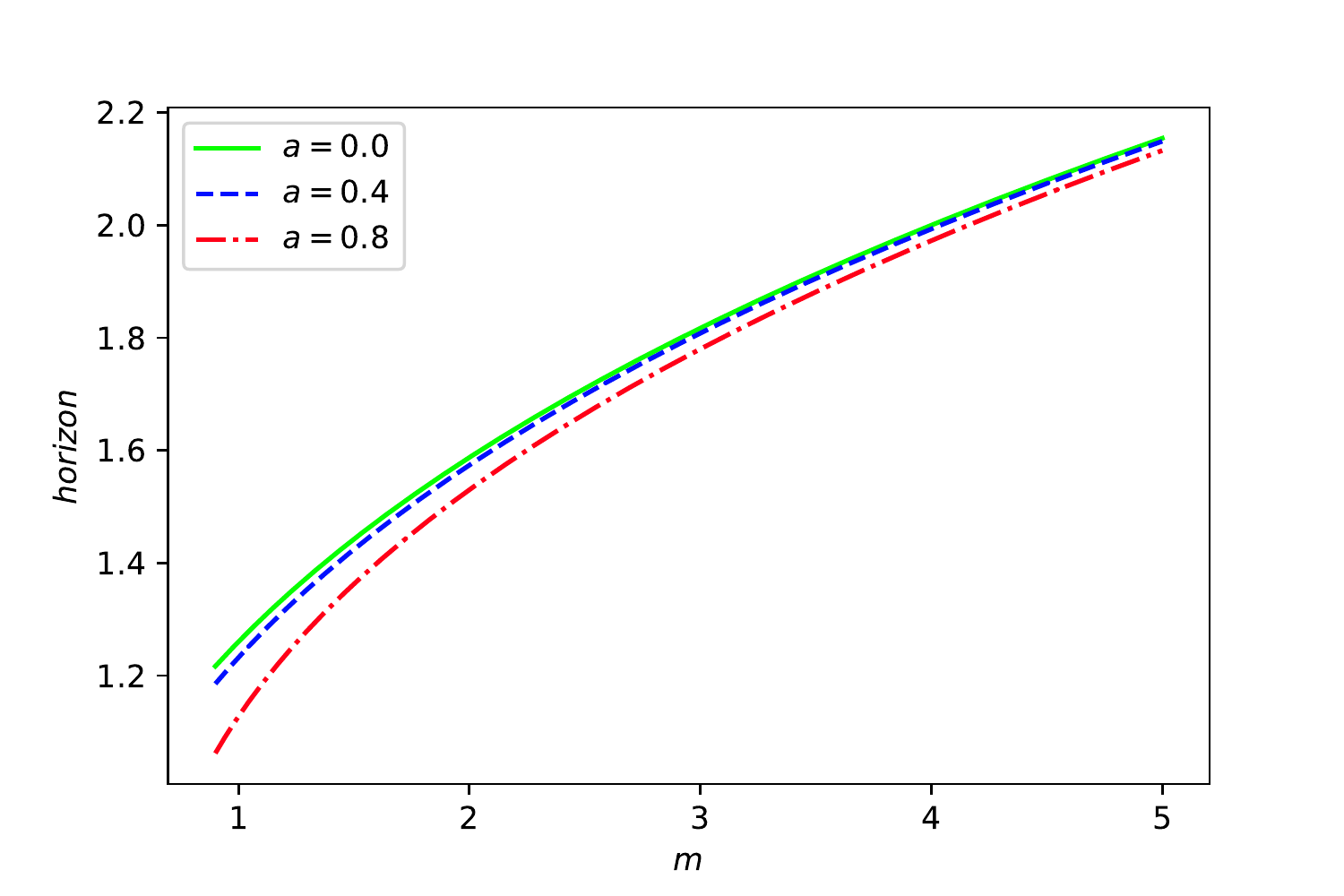}
	\caption{Mass and the horizon ($\mathbb T$): $d=4$}
	\label{fig:topBH_t_horizon}
	\end{minipage}
\end{figure}

\subsection{Spherical case}\label{subsec:topKABH_sphere}
The metric is, if there is no spherical term (we consider a case where the string moves around the equator parametrized by $\phi$), 
\begin{align}
ds_{\text{KA}\mathbb S}^2
&= -\frac{\Delta_r-\Delta_\theta a^2\sin^2\theta}{\rho^2}dt^2
  + \frac{\rho^2}{\Delta_r}dr^2
  + \frac{\rho^2}{\Delta_\theta}d\theta^2\nonumber\\
&\qquad
 + 2\frac{\Delta_r - \Delta_\theta(r^2+a^2)}{\rho^2\Xi}a\sin^2\theta dtd\phi
 + \frac{\Delta_\theta(r^2+a^2)^2 - \Delta_ra^2\sin^2\theta}{\rho^2\Xi^2}\sin^2\theta d\phi^2.
\end{align}
Taking the same assumption as before \eqref{eq:static_embedding}, the induced metric on worldsheet 
($t = \tau$, $r = r(\sigma)$, $\theta = \theta(\sigma)$ and $\phi = \phi(\sigma)$) is
\begin{align}
ds_{\text{KA}\mathbb S}^2|_\text{ind}
&= -\frac{\Delta_r-\Delta_\theta a^2\sin^2\theta}{\rho^2}d\tau^2\nonumber\\
&\quad  + \Big(\frac{\rho^2}{\Delta_r}r'^2
    + \frac{\rho^2}{\Delta_\theta}\theta'^2
    + \frac{\Delta_\theta(r^2+a^2)^2 - \Delta_ra^2\sin^2\theta}{\rho^2\Xi^2}
    \phi'^2\sin^2\theta\Big)d\sigma^2\nonumber\\
&\quad
  + 2\frac{\Delta_r - \Delta_\theta(r^2+a^2)}{\rho^2\Xi}a\phi'\sin^2\theta d\tau d\sigma.
\end{align}
Then the NG action is 
\begin{equation}
\frac1{T_\text{str}}\frac{dS_\text{NG}}{dt}
= \int d\sigma\sqrt{(\Delta_r-\Delta_\theta a^2\sin^2\theta)
  \Big(\frac{r'^2}{\Delta_r} + \frac{\theta'^2}{\Delta_\theta}\Big)
  + \frac{\Delta_r\Delta_\theta}{\Xi^2}\phi'^2\sin^2\theta},
\end{equation}
where as before the integral region is the inside of the horizon.
Changing the variables by $y=1/r$,
\begin{align}
\frac1{T_\text{str}}\frac{dS_\text{NG}}{dt}
&= \int d\sigma\sqrt{(\Delta_r-\Delta_\theta a^2\sin^2\theta)
  \Big(\frac{y'^2}{y^4\Delta_r} + \frac{\theta'^2}{\Delta_\theta}\Big)
  + \frac{\Delta_r\Delta_\theta}{\Xi^2}\phi'^2\sin^2\theta}.
\end{align}
For notational convenience, we define the following functions.
\begin{subequations}
\begin{align}
\Delta_y &:= y^4\Delta_{r=1/y} = (1+a^2y^2)(1+y^2) - 2my^d,\\
\Theta &:= \Delta_y - a^2y^4\sin^2\theta\Delta_\theta,\\
T &:= \frac{\Delta_y\Delta_\theta}{\Xi^2}\sin^2\theta.
\end{align}
\end{subequations}
The action becomes
\begin{equation}
\frac1{T_\text{str}}\frac{dS_\text{NG}}{dt}
= \int d\sigma \mathcal L_{\mathbb S},\;\;
\mathcal L_{\mathbb S} = \frac{L}{y^2},\;\;
L = \sqrt{\Theta\Big(\frac{y'^2}{\Delta_y} + \frac{\theta'^2}{\Delta_\theta}\Big) + T\phi'^2}.
\end{equation}

The equations of motion are 
\begin{subequations}
\begin{align}
&y'' - y'\frac{d}{d\sigma}\log L
 + y'\frac{d}{d\sigma}\log\Big(\frac{\Theta}{y^2\Delta_y}\Big)
 + \frac{\Delta_y}{\Theta}
  \Big(\frac{2L^2}{y} 
  - \frac12\frac{\partial L^2}{\partial y}\Big) = 0,\\
&\theta'' - \theta'\frac{d}{d\sigma}\log L
 + \theta'\frac{d}{d\sigma}\log\Big(\frac{\Theta}{y^2\Delta_\theta}\Big)
 - \frac{\Delta_\theta}{2\Theta}\frac{\partial L^2}{\partial\theta} = 0,\\
&\phi'' - \phi'\frac{d}{d\sigma}\log L
 + \phi'\frac{d}{d\sigma}\log\Big(\frac{T}{y^2}\Big) = 0.
\end{align}
\end{subequations}

We choose the gauge
$y'^2/\Delta_y + \theta'^2/\Delta_\theta = 1$
where $L = \sqrt{\Theta + T\phi'^2}$. 
Then the derivative of $L$ is
\begin{align}
L' &= \frac1{2L}(\Theta' + T'\phi'^2 + 2T\phi'\phi'')
= LA + \frac{T}{L}\phi'\phi'',\nonumber\\
A &:= \frac1{2L^2}(y'\partial\Delta_y - 4a^2y^3y'\Delta_\theta\sin^2\theta - a^2y^4\theta'\partial_\theta(\Delta_\theta\sin^2\theta) + T'\phi'^2).
\end{align}
In the matrix form the equations are summarized as
\begin{equation}
\begin{bmatrix}
1& & -Ty'\phi'/L^2\\ 
 & 1& -T\theta'\phi'/L^2\\
 & & 1-T\phi'^2/L^2
\end{bmatrix}
\begin{bmatrix}
y''\\ \theta''\\ \phi''
\end{bmatrix}
=
\begin{bmatrix}
y'A + y'B_y + C_y\\
\theta'A + \theta'B_\theta + C_\theta\\
\phi'A + \phi'B_\phi
\end{bmatrix},
\end{equation}
where we defined 
\begin{subequations}
\begin{align}
B_y &:= -\frac{d}{d\sigma}\log\Big(\frac{\Theta}{y^2\Delta_y}\Big),&
B_\theta &:= -\frac{d}{d\sigma}\log\Big(\frac{\Theta}{y^2\Delta_\theta}\Big),&
B_\phi &:= -\frac{d}{d\sigma}\log\Big(\frac{T}{y^2}\Big),\\
C_y &:= -\frac{\Delta_y}{\Theta}
  \Big(\frac{2L^2}{y} 
  - \frac12\frac{\partial L^2}{\partial y}\Big),&
C_\theta &:= \frac{\Delta_\theta}{2\Theta}\frac{\partial L^2}{\partial\theta}.
\end{align}
\end{subequations}
By multiplying the inverse matrix, the equation is solved for the second derivative terms:
\begin{equation}
\begin{bmatrix}
y''\\ \theta''\\ \phi''
\end{bmatrix}
=
\frac1{1-T\phi'^2/L^2}
\begin{bmatrix}
1-T\phi'^2/L^2& 0& Ty'\phi'/L^2\\
0& 1-T\phi'^2/L^2& T\theta'\phi'/L^2\\
0& 0& 1
\end{bmatrix}
\begin{bmatrix}
y'A + y'B_y + C_y\\
\theta'A + \theta'B_\theta + C_\theta\\
\phi'A + \phi'B_\phi
\end{bmatrix}.
\end{equation}
In the above the each factor is
\begin{subequations} 
\begin{align}
B_y &= \frac{2y'}y - \frac{\Theta'}{\Theta} 
 + \frac{y'\partial\Delta_y}{\Delta_y},\;\;
B_\theta = \frac{2y'}y - \frac{\Theta'}{\Theta} 
 + \frac{\theta'\partial\Delta_\theta}{\Delta_\theta},\;\;
B_\phi = \frac{2y'}y - \frac{T'}{T},\\
C_y &= 
-\frac{\Delta_y}{\Theta}
  \Big(\frac{2L^2}{y} 
  - \frac12\frac{\partial L^2}{\partial y}\Big),\;\;
C_\theta = 
\frac{\Delta_\theta}{2\Theta}\frac{\partial L^2}{\partial\theta},\\
\Theta' &= y'\partial\Delta_y - 4a^2y^3y'\sin^2\theta\Delta_\theta
 - a^2y^4\theta'\partial(\sin^2\theta\Delta_\theta),\\
\frac{T'}T &= \frac{y'\partial\Delta_y}{\Delta_y} 
 + \frac{\theta'\partial(\sin^2\theta\Delta_\theta)}{(\sin^2\theta\Delta_\theta)},\\
\partial\Delta_y 
&= 4a^2y^3 + 2(a^2+1)y - 2dmy^{d-1},\;\;
\partial\Delta_\theta
= a^2\sin(2\theta),\\
\partial(\sin^2\theta\Delta_\theta)
&= \sin(2\theta)(1-a^2\cos(2\theta)).
\end{align}
\end{subequations}
and 
\begin{subequations}
\begin{align}
\frac{\partial L^2}{\partial y}
&= \frac{\partial}{\partial y}(\Theta + T\phi'^2)
= \Big(1+\frac{\sin^2\theta\Delta_\theta}{\Xi^2}\phi'^2\Big)\partial\Delta_y
 - 4a^2y^3\sin^2\theta\Delta_\theta,\\
\frac{\partial L^2}{\partial\theta}
&= \frac{\partial}{\partial\theta}(\Theta + T\phi'^2)
= \Big(-a^2y^4+\frac{\Delta_y}{\Xi^2}\phi'^2\Big)\partial(\sin^2\theta\Delta_\theta).
\end{align}
\end{subequations}

\paragraph{Boundary condition}
The boundary condition is by the Neumann boundary condition $d\theta/dy = d\phi/dy = 0$  and by the gauge condition $y'^2/\Delta_y + \theta'^2/\Delta_\theta=1$ at $y=0$,
\begin{equation}
1 = y'^2 + \frac{\theta'^2}{1-a^2\cos^2\theta_0},\qquad
\therefore (y,\theta,\phi,y',\theta',\phi')
\overset{y\rightarrow0}\rightarrow(0,\theta_0,0,1,0,0).
\end{equation}

\begin{figure}[h]
	\begin{minipage}[t]{0.5\linewidth}
	\includegraphics[width=\linewidth]{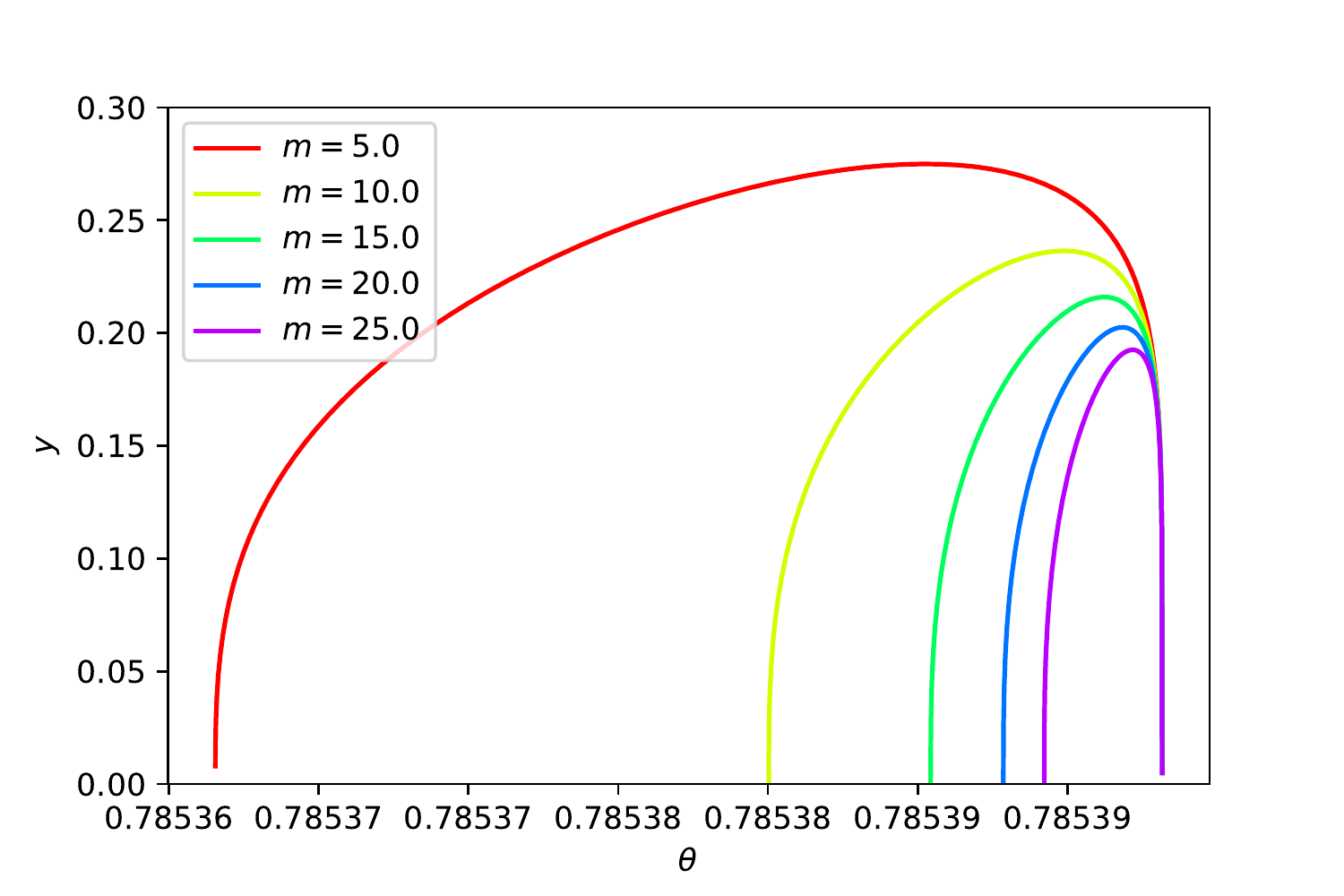}
	\caption{Mass dependence of the string embedding ($\mathbb S$, $\theta_0=\pi/4$)}
	\label{fig:topBHs_theta_y_m}
	\end{minipage}
\hspace{0\linewidth}
	\begin{minipage}[t]{0.5\linewidth}
	\includegraphics[width=\linewidth]{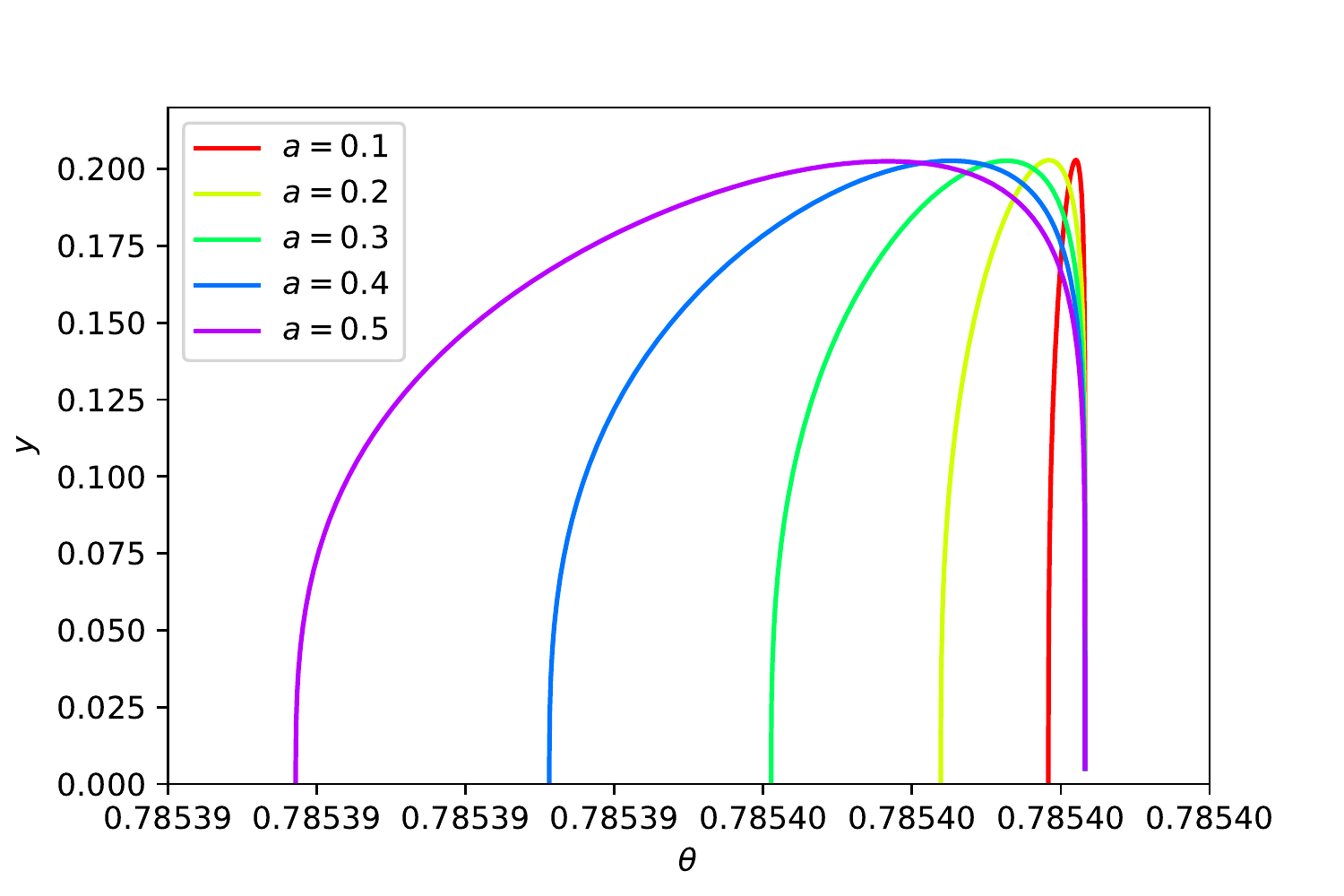}
	\caption{Angular momentum dependence of the string embedding ($\mathbb S$)}
	\label{fig:topBHs_theta_y_a}
	\end{minipage}
\hspace{0\linewidth}
	\begin{minipage}[t]{0.5\linewidth}
	\includegraphics[width=\linewidth]{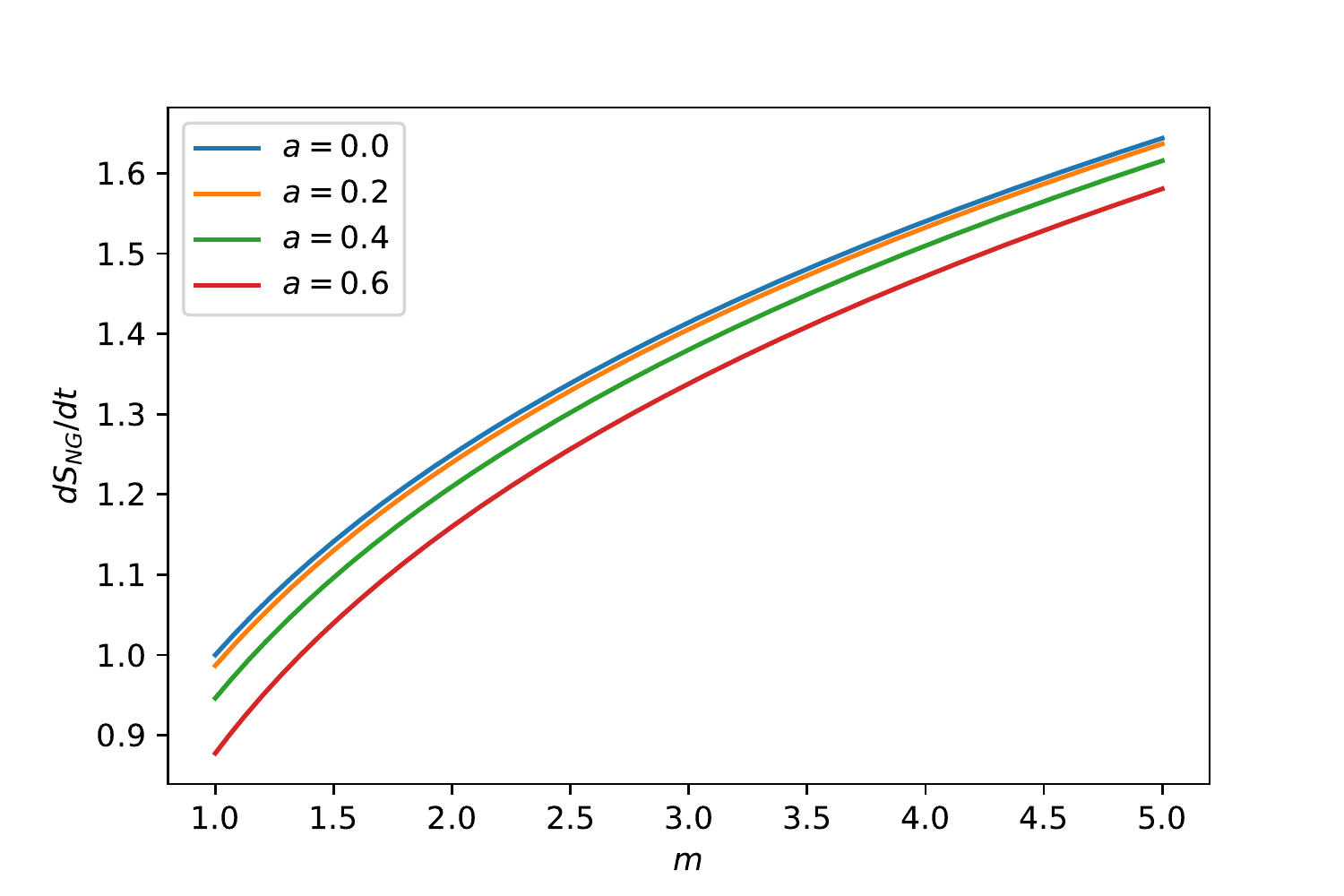}
	\caption{Mass and action growth for various angular momentums ($\mathbb S$)}
	\label{fig:topBH_s_m_action}
	\end{minipage}
\end{figure}

The results are plotted in 4-dimensional case.
Figure \ref{fig:topBHs_theta_y_m} represents the mass dependence for fixed angular momentum $a=0.5$.
The boundary condition for $\theta$ is $\theta(0) = \pi/4$.
We found the string does not penetrate the horizon in this case.
For example, $y_\text{h} = 0.62$ for mass $m=5$ and $y_\text{h} = 0.39$ for mass $m=25$.

The angular momentum dependence is plotted in Figure \ref{fig:topBHs_theta_y_a}.

\paragraph{WDW action}
For $\theta_0=0$ the action is 
\begin{equation}
\frac1{T_\text{str}}\frac{dS_\text{NG}}{dt}
= \int_{y_\text{h}}^\infty\frac{dy}{y^2}
= \frac1{y_\text{h}} =: x,
\end{equation}
where $y_\text{h}$ is determined by
\begin{equation}
\Delta_y(y_\text{h}) 
= (1+a^2y_\text{h}^2)(1+y_\text{h}^2) - 2my_\text{h}^d = 0.
\end{equation}
For $d=4$, by solving $x^4 + (1+a^2)x^2 - (2m-a^2) = 0$, ($x>0$),
\begin{equation}
\frac1{T_\text{str}}\frac{dS_\text{NG}}{dt}
= \bigg(\frac{-(1+a^2)+\sqrt{(1+a^2)^2+4(2m-a^2)}}2\bigg)^{1/2}.
\end{equation}
The mass dependence of the action are plotted in Figure \ref{fig:topBH_s_m_action}.

\subsection{Hyperbolic case}\label{subsec:topKABH_hyperbolic}
The metric is, if there is not spherical term, 
\begin{align}
ds_{\text{KA}\mathbb H}^2
&= -\frac{\Delta_r-\Delta_\theta a^2\sinh^2\theta}{\rho^2}dt^2
  + \frac{\rho^2}{\Delta_r}dr^2
  + \frac{\rho^2}{\Delta_\theta}d\theta^2
  - 2\frac{\Delta_r + \Delta_\theta(r^2+a^2)}{\rho^2\Xi}a\sinh^2\theta dtd\phi\nonumber\\
&\hspace{5cm}
 + \frac{\Delta_\theta(r^2+a^2)^2 - \Delta_ra^2\sinh^2\theta}{\rho^2\Xi^2}\sinh^2\theta d\phi^2.
\end{align}
On the worldsheet ($t=\tau, r=r(\sigma), \theta=\theta(\sigma)$ and $\phi=\phi(\sigma)$), the induced metric is 
\begin{align}
ds_{\text{KA}\mathbb H}^2|_\text{ind}
&= -\frac{\Delta_r-\Delta_\theta a^2\sinh^2\theta}{\rho^2}d\tau^2\nonumber\\
&\quad
  + \Big(\frac{\rho^2}{\Delta_r}r'^2
  + \frac{\rho^2}{\Delta_\theta}\theta'^2
  + \frac{\Delta_\theta(r^2+a^2)^2 - \Delta_ra^2\sinh^2\theta}{\rho^2\Xi^2}\sinh^2\theta\phi'^2\Big)d\sigma^2\nonumber\\
&\quad
  - 2\frac{\Delta_r + \Delta_\theta(r^2+a^2)}{\rho^2\Xi}a\phi'\sinh^2\theta d\tau d\sigma.
\end{align}
The Lagrangian is, by changing variables $y = 1/r$,
\begin{align}
\frac1{T_\text{str}}\frac{dS_\text{NG}}{dt}
&= \int d\sigma\sqrt{(\Delta_r - \Delta_\theta a^2\sinh^2\theta)
  \Big(\frac{r'^2}{\Delta_r} + \frac{\theta'^2}{\Delta_\theta}\Big)
  + \frac{\Delta_r\Delta_\theta}{\Xi^2}\phi'^2\sinh^2\theta}\nonumber\\
&= \int d\sigma\sqrt{(\Delta_r - \Delta_\theta a^2\sinh^2\theta)
  \Big(\frac1{y^4}\frac{y'^2}{\Delta_r} + \frac{\theta'^2}{\Delta_\theta}\Big)
  + \frac{\Delta_r\Delta_\theta}{\Xi^2}\phi'^2\sinh^2\theta}.
\end{align}
For later convenience, we define
\begin{align}
\frac1{T_\text{str}}\frac{dS_\text{NG}}{dt}
&= \int d\sigma\mathcal L_{\mathbb H},\;\;
\mathcal L_{\mathbb H} = \frac{L}{y^2},\nonumber\\
L &:= \sqrt{(\Delta_y-\Delta_\theta a^2y^4\sinh^2\theta)
  \Big(\frac{y'^2}{\Delta_y} + \frac{\theta'^2}{\Delta_\theta}\Big) + \frac{\Delta_y\Delta_\theta}{\Xi^2}\phi'^2\sinh^2\theta}.
\end{align}
In the same way as the spherical case, we define
\begin{subequations}
\begin{align}
L &= \sqrt{\Theta\Big(\frac{y'^2}{\Delta_y} + \frac{\theta'^2}{\Delta_\theta}\Big) + T\phi'^2},\\
&\Theta := \Delta_y - a^2y^4\sinh^2\theta\Delta_\theta,\;\;
T := \frac{\Delta_y\Delta_\theta}{\Xi^2}\sinh^2\theta,\\
&\Delta_y := (1+a^2y^2)(1-y^2) - 2my^d,\;\;
\Delta_\theta := 1+a\cosh^2\theta.
\end{align}
\end{subequations}
The equations of motion are
\begin{subequations}
\begin{align}
&y''- y'\frac{d}{d\sigma}\log L
 + y'\frac{d}{d\sigma}\log\Big(\frac{\Theta}{y^2\Delta_y}\Big)
 + \frac{\Delta_y}{\Theta}\Big(\frac{2L^2}{y}
  - \frac12\frac{\partial L^2}{\partial y}\Big) = 0,\\
&\theta''- \theta'\frac{d}{d\sigma}\log L 
 + \theta'\frac{d}{d\sigma}\log\Big(\frac{\Theta}{y^2\Delta_\theta}\Big)
 - \frac{\Delta_\theta}{2\Theta}
  \frac{\partial L^2}{\partial\theta} = 0,\\
&\phi''- \phi'\frac{d}{d\sigma}\log L
 + \phi'\frac{d}{d\sigma}\log\Big(\frac{T}{y^2}\Big) = 0.
\end{align}
\end{subequations}
We choose the gauge
$
y'^2/\Delta_y + \theta'^2/\Delta_\theta = 1,\;
L = \sqrt{\Theta + T\phi'^2}
$. 
Then the derivative of $L$ is
\begin{align}
L' &= LA + \frac{T}{L}\phi'\phi'',\\
A &:= \frac1{2L^2}(y'\partial\Delta_y - 4a^2y^3y'\Delta_\theta\sinh^2\theta 
	- a^2y^4\theta'\partial_\theta(\Delta_\theta\sinh^2\theta) + T'\phi'^2),
\end{align}
where we defined 
\begin{subequations}\label{eq:coef_hyperbolic}
\begin{align}
B_y &:= -\frac{d}{d\sigma}\log\Big(\frac{\Theta}{y^2\Delta_y}\Big),&
B_\theta &:= -\frac{d}{d\sigma}\log\Big(\frac{\Theta}{y^2\Delta_\theta}\Big),&
B_\phi &:= -\frac{d}{d\sigma}\log\Big(\frac{T}{y^2}\Big),\\
C_y &:= -\frac{\Delta_y}{\Theta}
  \Big(\frac{2L^2}{y} 
  - \frac12\frac{\partial L^2}{\partial y}\Big),&
C_\theta &:= \frac{\Delta_\theta}{2\Theta}\frac{\partial L^2}{\partial\theta}.
\end{align}
\end{subequations}
These are explicitly
\begin{equation}
B_y = \frac{2y'}y - \frac{\Theta'}{\Theta} 
 + \frac{y'\partial\Delta_y}{\Delta_y},\quad
B_\theta = \frac{2y'}y - \frac{\Theta'}{\Theta} 
 + \frac{\theta'\partial\Delta_\theta}{\Delta_\theta},\quad
B_\phi = \frac{2y'}y - \frac{T'}{T}.
\end{equation}
In the above
\begin{subequations}
\begin{align}
\Theta'
&= y'\partial\Delta_y - 4a^2y^3y'\sinh^2\theta\Delta_\theta
 - a^2y^4\theta'\partial(\sinh^2\theta\Delta_\theta),\\
\frac{T'}T 
&= \frac{y'\partial\Delta_y}{\Delta_y} 
 + \frac{\theta'\partial(\sinh^2\theta\Delta_\theta)}{(\sinh^2\theta\Delta_\theta)},\\
\partial\Delta_y 
&= -4a^2y^3 + 2(a^2-1)y - 2dmy^{d-1},\\
\partial\Delta_\theta
&= a^2\sinh(2\theta),\\
\partial(\sinh^2\theta\Delta_\theta)
&= \sinh(2\theta)(1+a^2\cosh(2\theta)),
\end{align}
\end{subequations}
and
\begin{subequations}
\begin{align}
\frac{\partial L^2}{\partial y}
&= \frac{\partial}{\partial y}(\Theta + T\phi'^2)
= \Big(1+\frac{\sinh^2\theta\Delta_\theta}{\Xi^2}\phi'^2\Big)\partial\Delta_y
 - 4a^2y^3\sinh^2\theta\Delta_\theta,\\
\frac{\partial L^2}{\partial\theta}
&= \frac{\partial}{\partial\theta}(\Theta + T\phi'^2)
= \Big(-a^2y^4+\frac{\Delta_y}{\Xi^2}\phi'^2\Big)
 \partial(\sinh^2\theta\Delta_\theta)
\end{align}
\end{subequations}
In the matrix form the equations are
\begin{equation}
\begin{bmatrix}
1& & -Ty'\phi'/L^2\\ 
 & 1& -T\theta'\phi'/L^2\\
 & & 1-T\phi'^2/L^2
\end{bmatrix}
\begin{bmatrix}
y''\\ \theta''\\ \phi''
\end{bmatrix}
=
\begin{bmatrix}
y'A + y'B_y + C_y\\
\theta'A + \theta'B_\theta + C_\theta\\
\phi'A + \phi'B_\phi
\end{bmatrix}.
\end{equation}
By multiplying the inverse matrix, these equations can be solved for the second derivative terms in the same way as the spherical case
\begin{equation}
\begin{bmatrix}
y''\\ \theta''\\ \phi''
\end{bmatrix}
=
\frac1{1-T\phi'^2/L^2}
\begin{bmatrix}
1-T\phi'^2/L^2& 0& Ty'\phi'/L^2\\
0& 1-T\phi'^2/L^2& T\theta'\phi'/L^2\\
0& 0& 1
\end{bmatrix}
\begin{bmatrix}
y'A + y'B_y + C_y\\
\theta'A + \theta'B_\theta + C_\theta\\
\phi'A + \phi'B_\phi
\end{bmatrix},
\end{equation}
where the coefficients are replaced with \eqref{eq:coef_hyperbolic}.
\paragraph{Boundary condition}
The boundary condition is by the Neumann boudary condition $d\theta/dy = d\phi/dy = 0$  and by the gauge condition $y'^2/\Delta_y + \theta'^2/\Delta_\theta=1$ at $y=0$,
\begin{equation}
1 = y'^2 + \frac{\theta'^2}{1+a^2\cosh^2\theta_0},\qquad
\therefore (y,\theta,\phi,y',\theta',\phi')
\overset{y\rightarrow0}\rightarrow(0,\theta_0,0,1,0,0).
\end{equation}

The results for 4-dimension are plotted in the following figures.
Figure \ref{fig:topBHh_theta_y_m} represents the mass dependence.
For example, the location of the horizon is $y_\text{h} = 0.53$ for $m=5$ and $y_\text{h} = 0.37$ for $m=25$.
The angular momentum dependence is plotted in Figure \ref{fig:topBHh_theta_y_a}.

\begin{figure}[h]
	\begin{minipage}[t]{0.5\linewidth}
	\includegraphics[width=\linewidth]{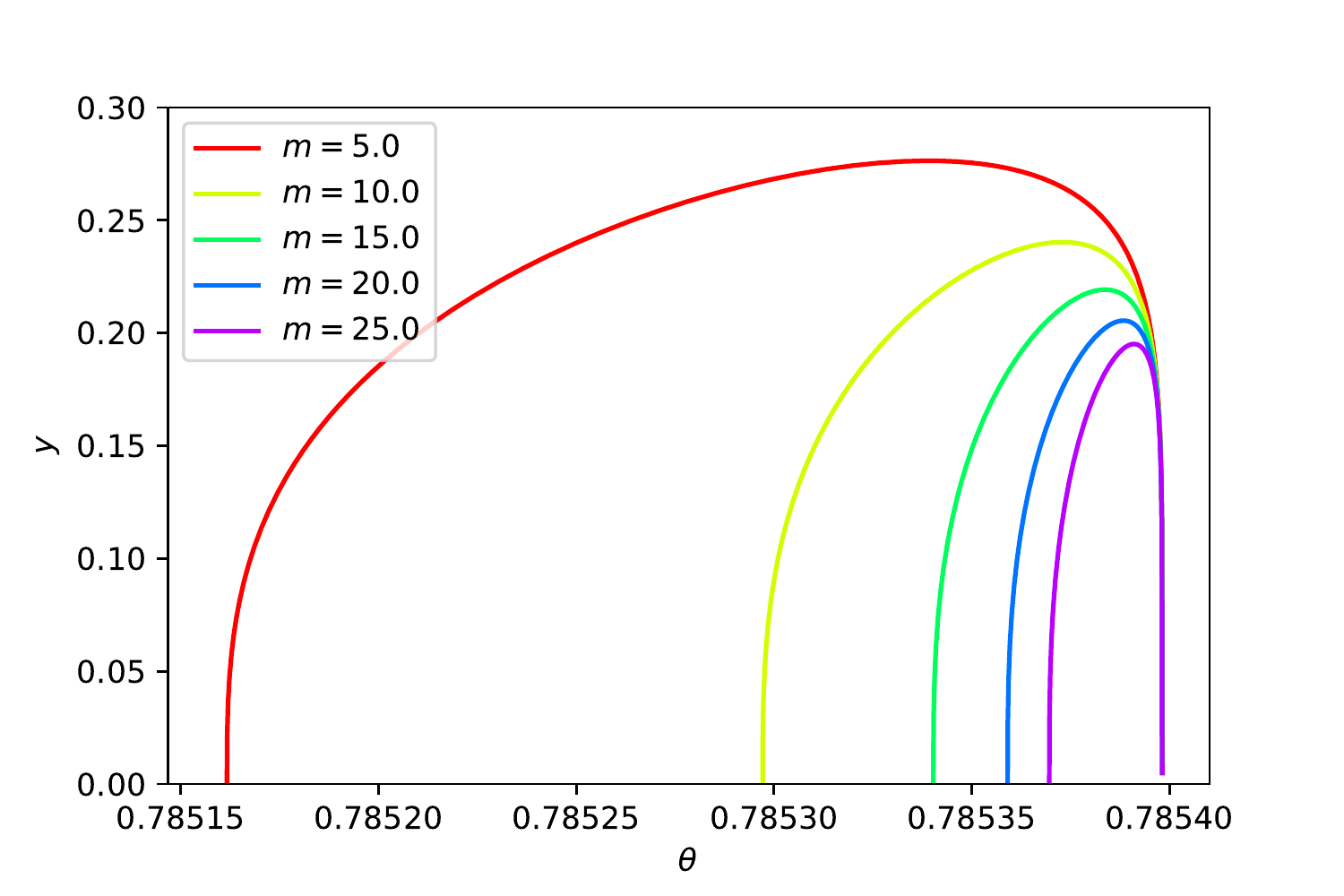}
	\caption{Mass dependence of the string embedding ($\mathbb H$, $\theta_0=\pi/4$)}
	\label{fig:topBHh_theta_y_m}
	\end{minipage}
\hspace{0\linewidth}
	\begin{minipage}[t]{0.5\linewidth}
	\includegraphics[width=\linewidth]{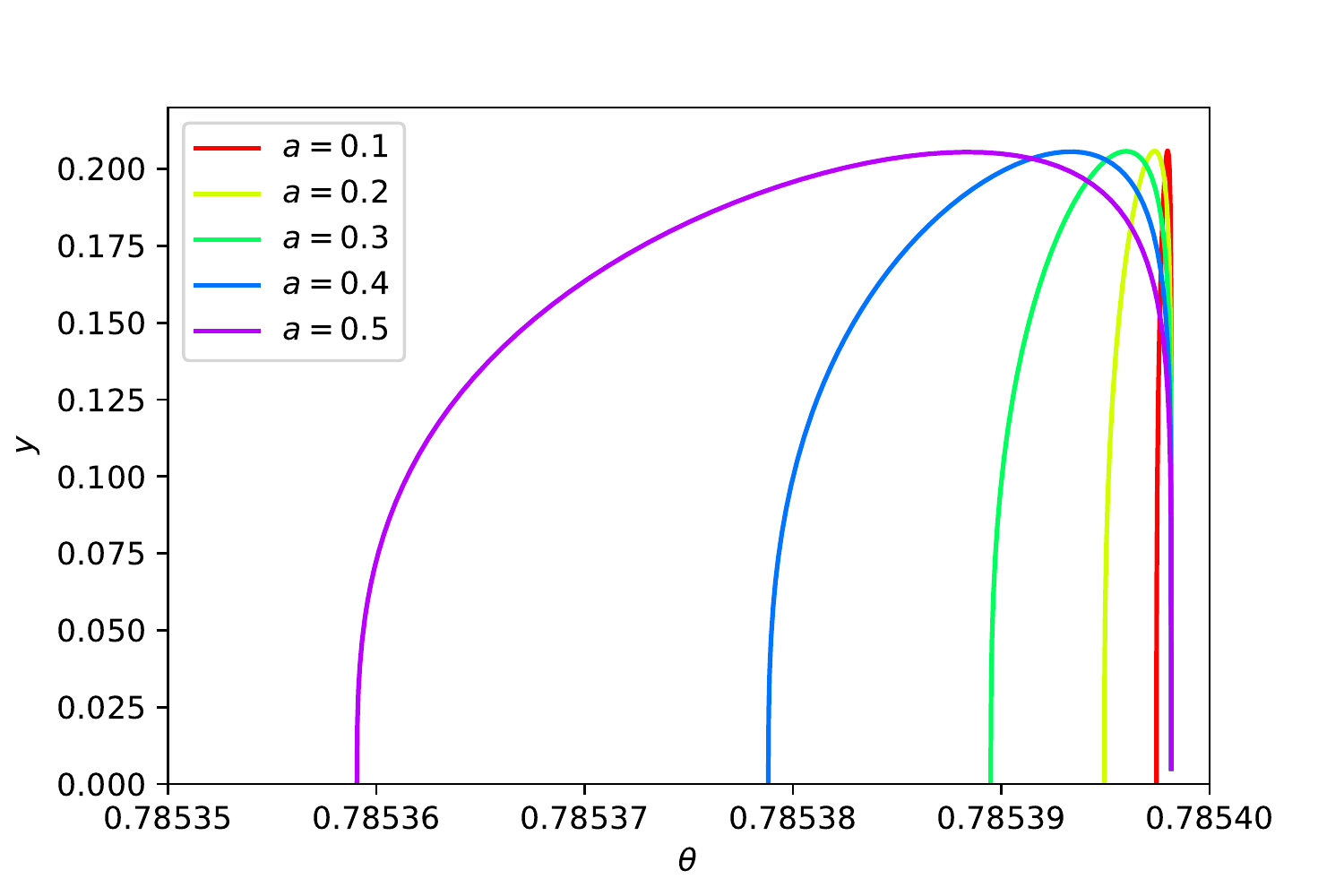}
	\caption{Angular momentum dependence of the string embedding ($\mathbb H$)}
	\label{fig:topBHh_theta_y_a}
	\end{minipage}
\hspace{0\linewidth}
	\begin{minipage}[t]{0.5\linewidth}
	\includegraphics[width=\linewidth]{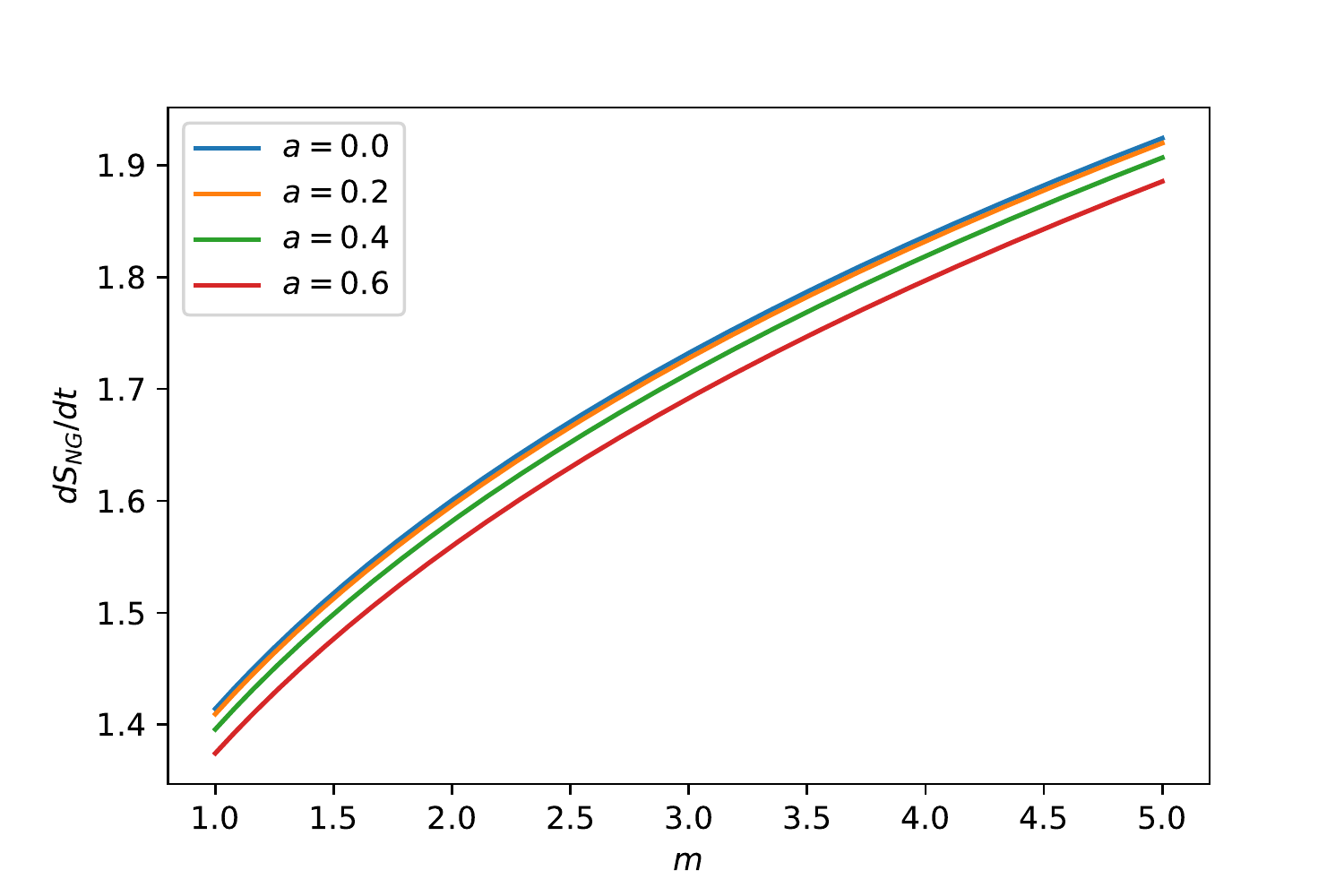}
	\caption{Mass and action growth for various angular momentums ($\mathbb H$)}
	\label{fig:topBH_h_m_action}
	\end{minipage}
\end{figure}

Comparing Figure \ref{fig:topBHs_theta_y_m} and Figure \ref{fig:topBHh_theta_y_m} for mass dependence and 
Figure \ref{fig:topBHs_theta_y_a} and Figure \ref{fig:topBHh_theta_y_a} for angular momentum dependence, we can see the string is more affected in the hyperbolic case. 

\paragraph{WDW action}
For $\theta_0=0$ the action is 
\begin{equation}
\frac1{T_\text{str}}\frac{dS_\text{NG}}{dt}
= \int_{y_\text{h}}^\infty\frac{dy}{y^2}
= \frac1{y_\text{h}} =: x,
\end{equation}
where $y_\text{h}$ is determined by
\begin{equation}
\Delta_y(y_\text{h}) 
= (1+a^2y_\text{h}^2)(1-y_\text{h}^2) - 2my_\text{h}^d = 0.
\end{equation}
For $d=4$, by solving $x^4 - (1-a^2)x^2 - (2m+a^2) = 0$, ($x>0$),
\begin{equation}
\frac1{T_\text{str}}\frac{dS_\text{NG}}{dt}
= \bigg(\frac{(1-a^2)+\sqrt{(1-a^2)^2+4(2m+a^2)}}2\bigg)^{1/2}.
\end{equation}
The mass dependence for different angular momentums are plotted in Figure 
\ref{fig:topBH_h_m_action}.

\subsection{Toroidal case}\label{subsec:topKABH_torus}
The induced metric of the toroidal rotating black hole is, if the spherical part is zero,
\begin{equation}
ds_{\text{KA}\mathbb T}^2|_\text{ind}
= -\Big(N^2 - \frac{\Sigma^2\omega^2}{\rho^2}\Big)d\tau^2 
  + \Big(\frac{\rho^2}{\Delta_r}r'^2
    + \frac{\rho^2}{\Delta_\theta}\theta'^2
    + \frac{\Sigma^2}{\rho^2}\phi'^2\Big)d\sigma^2
  - 2\frac{\omega\Sigma^2\phi'}{\rho^2}d\tau d\sigma.
\end{equation}
In the above we used
\begin{subequations}
\begin{align}
\rho^2 &:= r^2 + a^2\theta^2,\;\;
\Delta_\theta := 1 + a^2\theta^4,\;\;
\Delta_r := a^2 - 2m + r^4,\\
\Sigma^2 &:= r^4\Delta_\theta - a^2\theta^4\Delta_r,\;\;
\omega := a\frac{\Delta_r\theta^2 + r^2\Delta_\theta}{\Sigma^2},\;\;
N^2 = \frac{\rho^2\Delta_\theta\Delta_r}{\Sigma^2}.
\end{align}
\end{subequations}
The action is 
\begin{equation}
\frac1{T_\text{str}}\frac{dS_\text{NG}}{dt}
= \int d\sigma\mathcal L_{\mathbb T},\quad
\mathcal L_{\mathbb T}
= \sqrt{\Big(\frac{\rho^4}{\Sigma^2}\Delta_\theta
  - \frac{\omega^2\Sigma^2}{\Delta_r}\Big)r'^2  
  + \Big(\frac{\rho^4}{\Sigma^2}\Delta_r
  - \frac{\omega^2\Sigma^2}{\Delta_\theta}\Big)\theta'^2
    + \Delta_r\Delta_\theta\phi'^2}.
\end{equation}
where the first and the second terms have the common factor
\begin{equation}
\rho^4\Delta_r\Delta_\theta - \omega^2\Sigma^4
= (r^2+a^2\theta^2)^2\Delta_r\Delta_\theta
 - a^2(\theta^2\Delta_r + r^2\Delta_\theta)^2
= (\Delta_r - a^2\Delta_\theta)\Sigma^2.
\end{equation}
Therefore, 
\begin{equation}
\mathcal L_{\mathbb T} 
= \sqrt{\frac{\Delta_r - a^2\Delta_\theta}{\Delta_r}r'^2  
  + \frac{\Delta_r - a^2\Delta_\theta}{\Delta_\theta}\theta'^2
    + \Delta_r\Delta_\theta\phi'^2}.
\end{equation}
Changing variables by $y=1/r$,
\begin{subequations}
\begin{align}
\frac1{T_\text{str}}\frac{dS_\text{NG}}{dt}
&= \int d\sigma\mathcal L_{\mathbb T},\quad
\mathcal L_{\mathbb T} = \frac{L}{y^2},\\
L &:= \sqrt{(\Delta_y - a^2y^4\Delta_\theta)
  \Big(\frac{y'^2}{\Delta_y} + \frac{\theta'^2}{\Delta_\theta}\Big)
  + \Delta_y\Delta_\theta\phi'^2}.
\end{align}
\end{subequations}
It is convenience to define the following:
\begin{subequations}
\begin{align}
L &= \sqrt{\Theta\Big(\frac{y'^2}{\Delta_y} + \frac{\theta'^2}{\Delta_\theta}\Big)
  + T\phi'^2},\\
&\Theta = \Delta_y - a^2y^4\Delta_\theta,\;\;
T = \Delta_y\Delta_\theta,\\
&\Delta_\theta = 1+a^2\theta^4,\;\;
\Delta_y = a^2y^4 - 2my^3 + 1.
\end{align}
\end{subequations}
The equations of motion are 
\begin{subequations}
\begin{align}
&y''- y'\frac{d}{d\sigma}\log L
 + y'\frac{d}{d\sigma}\log\Big(\frac{\Theta}{y^2\Delta_y}\Big)
 + \frac{\Delta_y}{\Theta}\Big(\frac{2L^2}{y}
  - \frac12\frac{\partial L^2}{\partial y}\Big) = 0,\\
&\theta''- \theta'\frac{d}{d\sigma}\log L 
 + \theta'\frac{d}{d\sigma}\log\Big(\frac{\Theta}{y^2\Delta_\theta}\Big)
 - \frac{\Delta_\theta}{2\Theta}
  \frac{\partial L^2}{\partial\theta} = 0,\\
&\phi''- \phi'\frac{d}{d\sigma}\log L
 + \phi'\frac{d}{d\sigma}\log\Big(\frac{T}{y^2}\Big) = 0.
\end{align}
\end{subequations}
In the matrix notation the equations are the same form as before 
\begin{equation}
\begin{bmatrix}
y''\\ \theta''\\ \phi''
\end{bmatrix}
=
\frac1{1-T\phi'^2/L^2}
\begin{bmatrix}
1-T\phi'^2/L^2& 0& Ty'\phi'/L^2\\
0& 1-T\phi'^2/L^2& T\theta'\phi'/L^2\\
0& 0& 1
\end{bmatrix}
\begin{bmatrix}
y'A + y'B_y + C_y\\
\theta'A + \theta'B_\theta + C_\theta\\
\phi'A + \phi'B_\phi
\end{bmatrix},
\end{equation}
where the coefficients are replaced with
\begin{subequations}
\begin{align}
A &:= \frac1{2L^2}(y'\partial\Delta_y - 4a^2y^3y'\Delta_\theta 
  - a^2y^4\theta'\partial\Delta_\theta + T'\phi'^2),\\
B_y &:= \frac{2y'}y - \frac{\Theta'}{\Theta} 
 + \frac{y'\partial\Delta_y}{\Delta_y},\;\;
B_\theta := \frac{2y'}y - \frac{\Theta'}{\Theta} 
 + \frac{\theta'\partial\Delta_\theta}{\Delta_\theta},\;\;
B_\phi := \frac{2y'}y - \frac{T'}{T},\\
C_y &:= -\frac{\Delta_y}{\Theta}
  \Big(\frac{2L^2}{y} 
  - \frac12\frac{\partial L^2}{\partial y}\Big),\;\;
C_\theta := 
\frac{\Delta_\theta}{2\Theta}\frac{\partial L^2}{\partial\theta}.
\end{align}
\end{subequations}
In the above
\begin{subequations}
\begin{align}
\Theta' &= y'\partial\Delta_y - 4a^2y^3y'\Delta_\theta
 - a^2y^4\theta'\partial\Delta_\theta,\\
\frac{T'}T &= \frac{y'\partial\Delta_y}{\Delta_y} 
 + \frac{\theta'\partial\Delta_\theta}{\Delta_\theta},\\
\partial\Delta_y &= 4a^2y^3 - 6my^2,\\
\partial\Delta_\theta &= 4a^2\theta^3,
\end{align}
\end{subequations}
and 
\begin{subequations}
\begin{align}
\frac{\partial L^2}{\partial y}
&= \frac{\partial}{\partial y}(\Theta + T\phi'^2)
= (1+\Delta_\theta\phi'^2)\partial\Delta_y
 - 4a^2y^3\Delta_\theta,\\
\frac{\partial L^2}{\partial\theta}
&= \frac{\partial}{\partial\theta}(\Theta + T\phi'^2)
= (-a^2y^4+\Delta_y\phi'^2)\partial\Delta_\theta.
\end{align}
\end{subequations}

\paragraph{Boundary condition}
The boundary condition is by the Neumann boudary condition $d\theta/dy = d\phi/dy = 0$  and by the gauge condition $y'^2/\Delta_y + \theta'^2/\Delta_\theta=1$ at $y=0$,
\begin{equation}
1 = y'^2 + \frac{\theta'^2}{1+a^2\theta_0^4},\qquad
\therefore (y,\theta,\phi,y',\theta',\phi')
\overset{y\rightarrow0}\rightarrow(0,\theta_0,0,1,0,0).
\end{equation}

\begin{figure}[h]
	\begin{minipage}[t]{0.5\linewidth}
	\includegraphics[width=\linewidth]{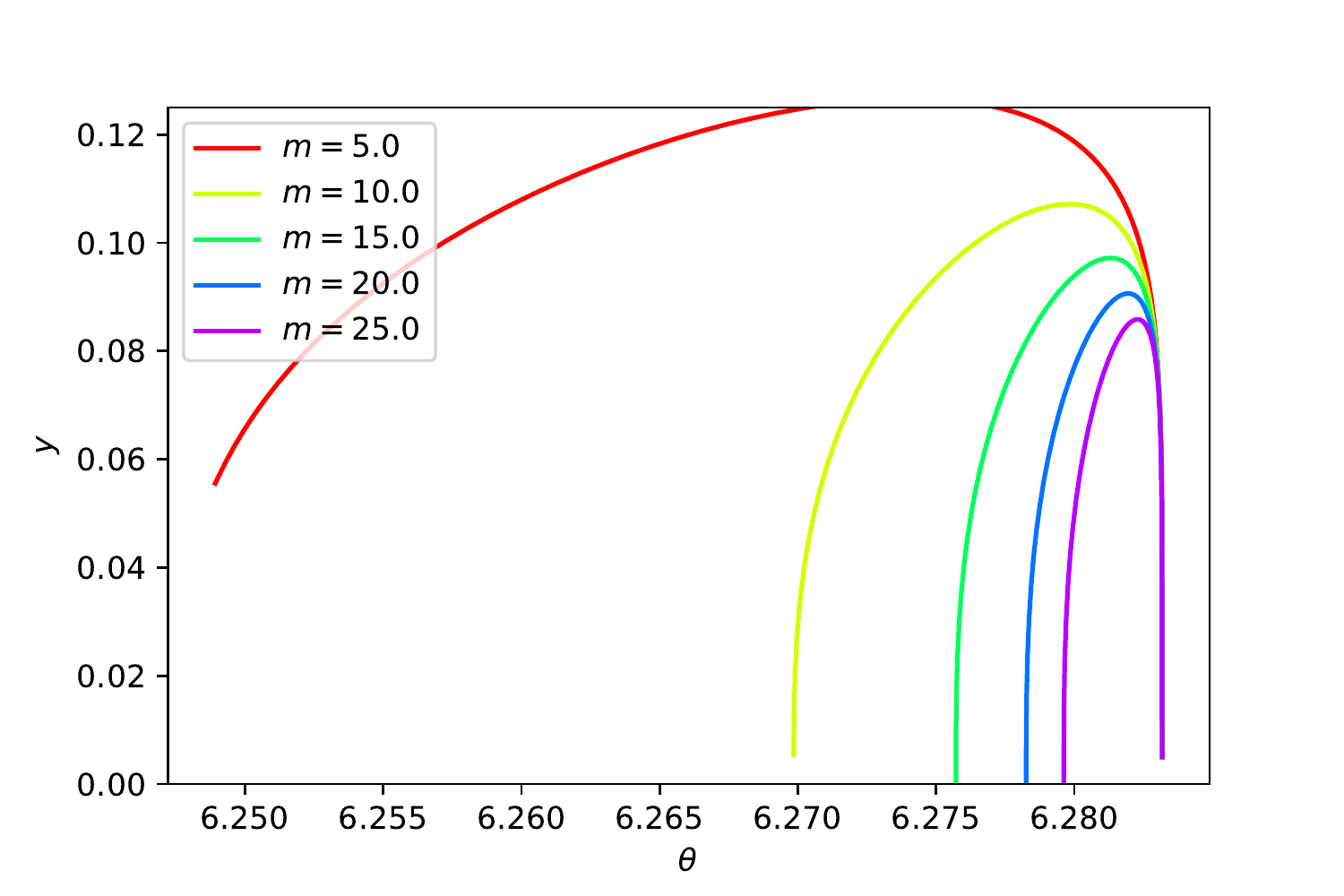}
	\caption{Mass dependence of the string embedding ($\mathbb T$)}
	\label{fig:topBHt_theta_y_m}
	\end{minipage}
\hspace{0\linewidth}
	\begin{minipage}[t]{0.5\linewidth}
	\includegraphics[width=\linewidth]{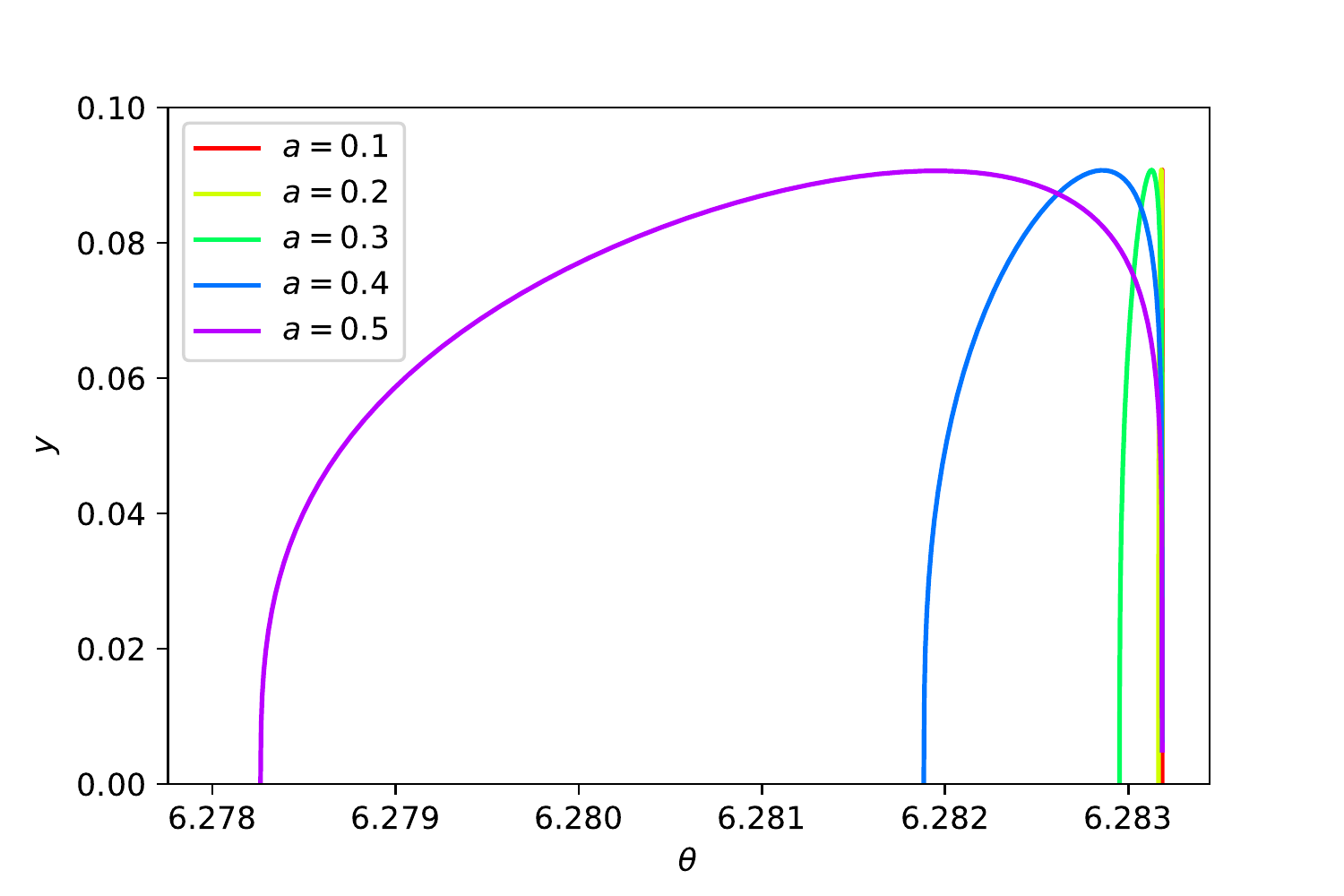}
	\caption{Angular momentum dependence of the string embedding ($\mathbb T$)}
	\label{fig:topBHt_theta_y_a}
	\end{minipage}
\end{figure}

Figure \ref{fig:topBHt_theta_y_m} represents the mass dependence. 
For example, $y_\text{h} = 0.47$ for $m=5$ and $y_\text{h} = 0.27$ for $m=25$.
The angular momentum dependence is plotted in Figure \ref{fig:topBHt_theta_y_a}.

For $\theta_0=0$ the action growth is 
\begin{equation}
\frac{dS_\text{NG}}{dt}
= \frac1{2\pi\alpha'}
 \int_{y_\text{h}}^\infty\frac{dy}{y^2}
 \sqrt{1-\frac{a^2y^4}{\Delta_y}}
= \frac1{2\pi\alpha'}
 \int_{y_\text{h}}^\infty\frac{dy}{y^2}
 \sqrt\frac{1-2my^3}{1-2my^3+a^2y^4}.
\end{equation}
In the square root the numerator is less than the denominator for non zero $a$.
Then it cannot penetrate the horizon for $a\neq0$.

Comparing Figure \ref{fig:topBHs_theta_y_m} and Figure \ref{fig:topBHt_theta_y_m} for mass dependence and 
Figure \ref{fig:topBHs_theta_y_a} and Figure \ref{fig:topBHt_theta_y_a} for angular momentum dependence, we can see the string is tend to receive more effect than sphere and hyperbolic cases as seen for small masses or large angular momentum.

\section{Discussion}\label{sec:Discussion}
In this paper we found the behavior of the fundamental strings in black hole spacetime with three different horizon structures.
In \eqref{eq:static_adsmetric} the curvature parameter $k$ distinguishes these cases: 
$k=1$ is the spherical horizon case, $k=-1$ is the hyperbolic case and $k=0$ is the toroidal horizon case.

First in Section \ref{sec:statictopBH} we studied the static black holes and found that the constant solution $\theta'=0$ is only able to approach the horizon for all cases ($k=\pm1,0$).
The growth of the Nambu-Goto (NG) action on the Wheeler DeWitt (WDW) is proportional to the horizon radius.
Only for the hyperbolic black holes, there is a non-trivial contribution even for zero-mass, where the metric function $g_{-1}(r)$ has a zero point $r_\text{h} = 1$.

\begin{figure}[t]
\begin{center}
	\includegraphics[width=15cm]{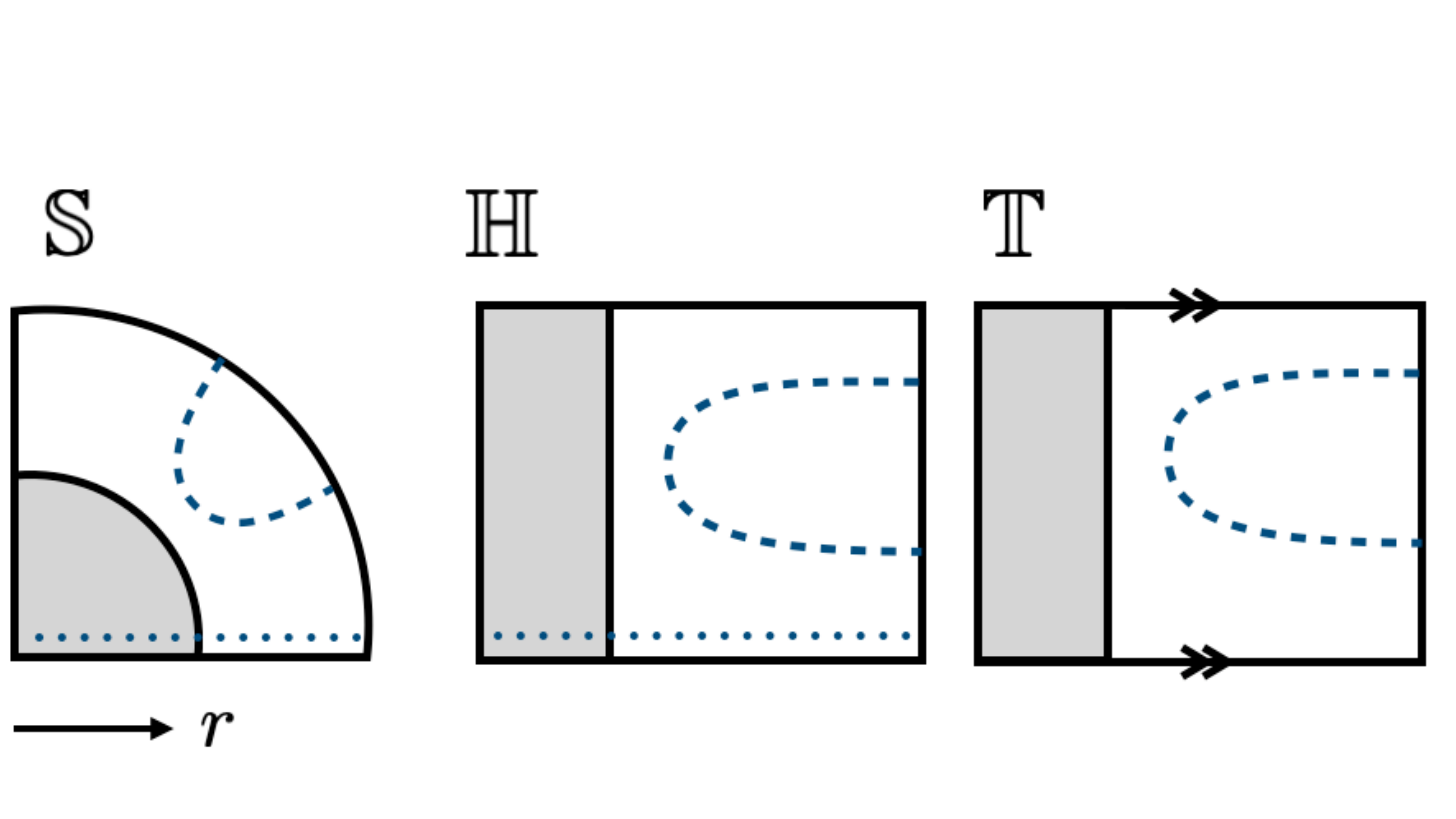}
	\caption{Sketch of string behavior in black hole spacetime for different horizon structures: 
	For each case the shaded region is the inside of the horizon. 
	The horizontal direction is the radial direction $r$ and the vertical direction is $\theta$.
	The right edge or the outer edge is the AdS boundary for each case.
	The dotted line which starts with $\theta = 0$ is the only case where the string can penetrates the horizon. 
	The dashed line starts with $\theta\neq0$ and such a string does not enter the horizon.
		Left: For spherical case ($\mathbb S$) the region of $\theta$ is $0\leq\theta\leq\pi$.
		Middle: For hyperbolic case ($\mathbb H$) the region of $\theta$ is $0\leq\theta$.
		Right: For Toroidal case ($\mathbb T$) the region of $\theta$ is $0\leq\theta\leq1$ and the top side and the bottom side are identified.}
\label{fig:sketch_topBH_string}
\end{center}
\end{figure}

For rotating topological black holes, the behavior of the strings are summarized in Figure \ref{fig:sketch_topBH_string}.
In this figure, the dashed lines represent the strings which bend and do not extend to the interior of the horizon.
As we found in the calculation, such strings approach the AdS boundary at non-zero altitude angle $\theta\neq0$ for the spherical and the hyperbolic cases.
In the left and the middle panels the dotted lines are strings which attach the boundary at $\theta = 0$.
It is the only case where the string can penetrate the horizon of the black holes.
A remarkable point is that there is not such a solution for toroidal case (the right panel).
For the toroidal case, the string can penetrate the horizon only if the black hole is not rotating.

For spherical $k=1$ and hyperbolic $k=-1$ cases, there exist solutions which penetrate the horizon if the strings attach the boundary at $\theta_0 = 0$.
In these cases we obtained the WDW action.
Figure \ref{fig:topBH_s_m_action} and Figure \ref{fig:topBH_h_m_action} say the growth of the action is an increasing function of mass and the angular momentum has the negative contribution to it.

For toroidal case, $k=0$, we found that there exists the string solution which penetrates the horizon only for zero angular momentum $a=0$.
In this case the effect of the fundamental string to the growth of the action is given in Figure \ref{fig:topBH_dim3_m} and Figure \ref{fig:topBH_dim4_m} ($k=0$) in Section \ref{sec:statictopBH}.
This is also an increasing function of mass. 
However, as mentioned in Introduction and the begging of Section \ref{sec:statictopBH_action}, in the static situation this does not have an interpretation for the drag force. 
Moreover, since the string cannot exist in the horizon in this case, the string is not a good probe to capture the property of complexity growth for toroidal black holes. 

Let us see the holographic interpretation of these spacetimes.
In the holographic gauge theory, the fundamental string we considered in this paper is interpreted as a non-local gauge theory object, ``a Wilson loop."
The edge of the string at the AdS boundary is a test particle with infinite mass.
Therefore we conclude that for the spherical and the hyperbolic horizon black holes there is a possibility that there are two particles or one particle on the AdS boundary while for the toroidal horizon case, there are always two particles on the dual gauge theory which are edges of one fundamental string.

Let us also discuss a condition which any physical system must satisfy.
For the rate of computation the system may satisfy ``Lloyd's bound," \cite{Lloyd:2000nature,
MARGOLUS1998188,PhysRevLett.90.167902}:
\begin{equation}\label{eq:Lloydbnd}
\frac{dS}{dt}\leq\frac2{\pi\hbar}(M-\Omega J),
\end{equation}
where $\Omega$ is the angular velocity and $J$ is the angular momentum. 
On the lefthand side of \eqref{eq:Lloydbnd}, $S$ is the total action which now consists of 
$S = S_\text{bulk} + S_\text{NG}$, where $S_\text{bulk}$ includes the Einstein-Hilbert term and the some boundary terms.
On the righthand side of the inequality \eqref{eq:Lloydbnd} $M$ is the total mass or energy of the system. 
In our case, not only the black hole mass, we have to take into account the mass of the string.
However, the string stretches towards the infinity of the space and it has the infinite length.
Therefore in our calculation this inequality is automatically satisfied. 

Finally from the result of this paper we can say that the angular momentum has the effect for decreasing the rate of the complexity growth for all cases ($\mathbb S$, $\mathbb H$ and $\mathbb T$).
This outcome is consistent with the above comment for Lloyd's bound which says the the upper bound decreases for rotating black holes. 
It may be counterintuitive that a dynamical system has the less growth rate of complexity but the same phenomena was also found in \cite{Nagasaki:2017kqe,Nagasaki:2018csh}, where we studied a moving fundamental string in AdS black hole spacetime. 
Furthermore, the growth rate of the complexity was smaller when the relative velocity between the string and the angular momentum of the black holes are small. 

\section*{Acknnowledgments}
I would like to thank Sung-Soo Kim and Satoshi Yamaguchi for helpful discussion.


\providecommand{\href}[2]{#2}\begingroup\raggedright\endgroup

\end{document}